\documentstyle[prl,aps,psfig]{revtex}
\begin{document} 
\draft
\title{The Haldane bosonisation scheme and metallic states of
interacting fermions in $d$ spatial dimensions
} 
\author{\sc Behnam Farid}
\address{Cavendish Laboratory, Department of Physics, 
University of Cambridge,\\
Cambridge CB3 0HE, United Kingdom 
\thanks{Electronic address: bf10000@phy.cam.ac.uk}\\
and Max-Planck-Institut f\"ur Festk\"orperforschung,
Heisenbergstra\ss e 1,\\
70569 Stuttgart, Federal Republic of Germany
\thanks{Electronic address: farid@audrey.mpi-stuttgart.mpg.de}\\ 
}
\date{\today }
\maketitle
\vspace{0.66cm}
\begin{abstract}
{\sf
We consider the Haldane bosonisation scheme in $d$ spatial dimensions
as applied to a realistic model of interacting fermions in $d=2$ and 
unequivocally demonstrate failure of this scheme in $d > 1$, specifically 
in $d=2$. In addition to tracing back this failure to its origin, we 
show that {\sl nothing} as regards the true metallic state of the model 
under consideration is known with any degree of certainty.
}
\end{abstract}
%\vspace{0.33cm}

%_____________________________________________________
\def\Bf#1{\mbox{\boldmath{$#1$}}}
%_____________________________________________________
%_________
\vskip 2.8cm
\noindent\underline{\sf To appear in:
{\sl Philosophical Magazine, Part B, Vol.~{\Bf 80}, No.~9 (2000)} }
\vskip 0.25cm
%_________

%\maketitle
%\vfill
%\pagebreak

% I.
\section{Introduction}

{\Large\bf P}roperties of high-temperature (hereafter hight-$T_c$) cuprate 
superconductors in their normal states indicate these to be non-Fermi 
liquids (NFLs) (Anderson 1997), yet after one decade of intensive efforts 
by researchers, correspondence of a NFL metallic state to a microscopic 
model (without magnetic impurities) in spatial dimensions $d$ greater than 
one has proved elusive: barring non-uniform and superconducting states (at 
absolute zero of temperature), these efforts have consistently borne out 
the existence of solely Fermi-liquid (FL) metallic states in $d > 1$ (for 
a review see Farid 1999a). In consequence, a specific category of NFLs, 
namely that of Luttinger's, has prevailed to be considered viable only in 
$d=1$. The existence of NFLs, specifically Luttinger liquids (LLs) (Haldane 
1980, 1981) in $d>1$, as postulated by Anderson (1988, 1989, 1990a,b, 1992, 
1993; Anderson and Ren 1990), not only would, according to Anderson and 
co-workers, provide the basis for a unified explanation of the anomalous 
normal-state properties of the high-$T_c$ cuprates, but also would account 
for the superconductivity-pairing mechanism in these materials (Wheatley, Hsu, 
Anderson 1988a,b, Anderson 1991a,b, 1992, Chakravarty, Sudb{\o}, Anderson 
and Strong 1993, Anderson 1997). On the basis of their theoretical studies 
--- which rely upon Haldane's (1992) non-perturbative bosonisation scheme 
for gap-less degrees of freedom of interacting fermions in arbitrary 
spatial dimensions $d$ --- of a microscopic model of interacting electrons 
in $d=2$, originally put forward by Houghton and Marston (1993), Houghton, 
Kwon and Marston (HKM) (1994) and Houghton, Kwon, Marston and Shankar 
(HKMS) (1994) have arrived at the conclusion that (see also Kwon, Houghton 
and Marston 1995) i) for fermions interacting through short-range and the 
long-range Coulomb interaction function, the metallic state in $d=2$ is a 
FL and that, in conformity with Bares and Wen (1993), ii) for super 
long-range interacting fermions the metallic state is a NFL. 
\footnote{\label{f1}
\small
To be precise, Bares and Wen (1993) refer to this state as a ``$Z_{k_F}=0$ 
Fermi liquid'', where $Z_{k_F}$ stands for the value of the jump in the 
momentum distribution function ${\sf n}(k)$ at Fermi momentum $k_F$. } 
In this work we unequivocally demonstrate that conclusions (i) and (ii) 
are biased in consequence of insufficiently detailed considerations and
application of an unjustified approximation employed by HKM (1994) and 
HKMS (1994). More importantly, we demonstrate that in its present form
the Haldane (1992) bosonisation method suffers from a fundamental 
shortcoming in $d > 1$, rendering any conclusion obtained through its 
application indefensible. We thus conclude that on the basis of the 
available knowledge, the true metallic state of the model under 
consideration {\sl cannot} be considered to be known with any degree of 
certainty.

The organisation of this work is as follows. In Section II we deduce some
leading terms in the asymptotic series of $\delta G(\omega)$, the difference 
between the interacting and non-interacting Green functions, for small 
values of energy $\omega$. Here we explicitly deal with the conventional 
FLs in two and three spatial dimensions. In Section III we briefly describe 
some aspects of the Haldane bosonisation scheme that are relevant to our 
considerations in this work and obtain, directly from the energy-momentum 
($\omega,{\Bf k}$) representation of $G$ within this scheme, the coefficient 
of the $1/\omega$ term in the asymptotic series for small $\omega$ of 
$\delta G(\omega)$. The sign of this coefficient turns out to be negative 
which is opposite to that corresponding to conventional FLs. Here we establish 
that this property is common to a class of un-conventional FLs where the 
Fermi energy of the interacting system coincides with that of the 
non-interacting system. Our subsequent analyses bring out that in fact 
according to the Haldane bosonisation scheme, {\sl all} metallic 
systems in $d > 1$ must belong to this category of unconventional FLs
(that is, this holds true independent of the nature and strength of the 
interaction amongst the fermions). We devote Section IV to a discussion of 
the long-distance behaviour of the advanced part of the single-particle 
Green function, that is $G({\Bf x}, t=0^-)$. Here we expose a close 
relationship that exists between this behaviour and some smoothness 
property of the momentum-distribution function ${\sf n}({\Bf k})$ and 
specifically determine the condition to be met by ${\sf n}({\Bf k})$ in 
order for the anomalous exponent of $G({\Bf x},t=0^-)$ to be vanishing, 
that is one that renders the large-$\|{\Bf x}\|$ behaviour of $G({\Bf x},
t=0^-)$ to be FL-like. Our finding in this section with regard to the 
just-indicated FL-like behaviour of $G({\Bf x},t=0^-)$ at large 
$\|{\Bf x}\|$ within the Haldane bosonisation scheme for $d > 1$, leads us 
to the conclusion that this condition must be built into the structure of 
this scheme. Consequently, in Section V we consider the algebraic structure 
underlying the Haldane programme and establish that the Kac-Moody algebra 
concerning some current operators, which is basic to this structure, 
neglects in $d > 1$ some contributions that are the more significant the 
greater the deviation from unity is of the amount of discontinuity in 
${\sf n}({\Bf k})$ at the Fermi momentum. In other words, in $d > 1$ the 
just-indicated Kac-Moody algebra turns out to give a correct account of 
the physical processes in the system {\sl only} when the fermions are 
strictly non-interacting. This finding clarifies the reason underlying 
our two observations mentioned above, namely that within the framework of 
the Haldane bosonisation scheme the Fermi energy of the interacting system 
coincides with that of the {\sl non-interacting} system and that the 
anomalous dimension corresponding to the large-$\|{\Bf x}\|$ behaviour of 
$G({\Bf x},t=0^-)$ is vanishing, a property that interacting and 
non-interacting FLs share. In Section VI we present a brief description 
of the existing alternative realisations of the original Haldane bosonisation 
approach in $d > 1$ and argue that none of them is capable of describing 
NFL metallic states in $d >1$. We conclude our work by a summary and 
some remarks.

% II.
\section{Preliminaries}

Classification of metallic states is easiest accomplished through that 
of the self-energy $\Sigma$ in the energy-momentum representation, for 
$\omega$ and ${\Bf k}$ close to the Fermi energy and Fermi surface (FS), 
respectively. For simplicity, consider a uniform system of spin-less 
fermions in $d$ spatial dimensions interacting through the isotropic 
two-body potential $v$. For $k$ and $\omega$ in the vicinities of the 
Fermi wave number $k_F$ and Fermi energy $\omega_F$ (both of which we 
identify with zero when no confusion can arise), respectively, FLs 
(un-conventional ones included) comprise those metallic states whose 
corresponding $\Sigma(k,0)$ and $\Sigma(0,\omega)$ are 
continuously-differentiable functions of $k$ and $\omega$ in neighbourhoods 
of $k=0$ and $\omega=0$, respectively (for details see Farid 1999a). 
In $d=3$, self-energies of {\sl conventional} FLs for $\omega\to 0$ and 
small $k$ asymptotically behave like 
\footnote{\label{f1a}
\small
We point out that here and in what follows we restrict our considerations
to the leading $\omega$-dependent asymptotic contributions to 
${\rm Re}\Sigma(k,\omega)$ and ${\rm Im}\Sigma(k,\omega)$, for
$\omega\to 0$, taken as independent functions; otherwise, taking into
account, for instance, $-i \alpha_k\, {\rm sgn}(\omega)\, \omega^2$,  
would require to take into account the {\sl real} contribution
$\gamma_k\, \omega^2 \ln\vert\omega\vert$ which is asymptotically
more dominant than $i\,\alpha_k\, {\rm sgn}(\omega) \,\omega^2$.}
$\Sigma(k,\omega)\sim\Sigma(k,0) + \beta_k\,\omega - i\,\alpha_k\, 
{\rm sgn}(\omega) \,\omega^2$ (Hugenholtz 1957, DuBois 1959, Luttinger 
1961) with $\alpha_k \ge 0$ (for $\beta_k$ see below). The same expression 
holds for {\sl conventional} FLs in $d=2$ when $k\not=0$; when $k=0$, 
however, $\omega^2$ in this expression is to be replaced by 
$-\omega^2\ln\vert\omega\vert$ (Hodges, Smith and Wilkins 
1971, Bloom 1975, Fujimoto 1990, Fukuyama, Narikio and Hasegawa 1991). 
It is for our further considerations relevant to mention that {\sl all}
FLs have in common that ${\rm Im}\Sigma(k,\omega) = o(\omega)$, where 
$f(\omega)= o\big(g(\omega)\big)$ signifies that $f(\omega)/g(\omega)
\to 0$ for $\omega\to 0$; thus $\vert\omega\vert^{1+\alpha}$ with 
$\alpha > 0$ qualifies to be denoted by $o(\omega)$. In consequence of 
this property, making use of the Kramers-Kr\"onig relation, i.e. the one 
expressing ${\rm Re}\Sigma(k,\omega)$ in terms of a principal-value 
integral of ${\rm Im}\Sigma(k,\omega)$, while invoking the requirement 
of stability of the {\sl interacting} system, it can be rigorously proved 
(Farid 1999b) that in the asymptotic series expansion of $\Sigma(k,
\omega)$, for $\omega\to 0$, pertaining to an {\sl interacting} FL 
system, a term must occur of the form $\beta_k\,\omega$ ({\sl c.f.} the 
above-presented expression) where $\beta_k < 0$ (strictly negative). It 
follows that an {\sl interacting} system for which $\beta_k=0$ for {\sl 
any} value of $k$ (with $k$ close to zero so that a separation of 
$\Sigma(k,\omega)$ into separate functions of $k$ and $\omega$ can be 
rigorously effected), {\sl cannot} be a FL.

The self-energy $\Sigma$ is related to the single-particle Green function 
$G$ through the Dyson equation (Fetter and Walecka 1971), which in the 
$\omega,k$-representation reads $G(k,\omega) = G_0(k,\omega) + \delta 
G(k,\omega)$, where $\delta G(k,\omega) {:=} G_0(k,\omega)\Sigma(k,\omega) 
G(k,\omega)$, with $G_0(k,\omega)$ and $G(k,\omega)$ the single-particle 
Green functions pertaining to the non-interacting and interacting systems, 
respectively. From this equation the following expression is readily 
obtained
\begin{equation}
\label{e1}
\delta G(k,\omega) =\frac{\Sigma(k,\omega)}{[G_0^{-1}(k,\omega)
-\Sigma(k,\omega)] G_0^{-1}(k,\omega)}. 
\end{equation}
In what follows for simplicity, but without loss of generality, we often 
consider the case corresponding to $k=0$. Consequently, when $k=0$, we 
suppress $k$ in the arguments of the pertinent functions; thus for 
instance $\Sigma(\omega)$ will denote $\Sigma(0,\omega)$. We note that 
for {\sl conventional} FLs holds $\Sigma(0)\not=0$.
\footnote{\label{f2}
\small
In the event that $\Sigma(0)=0$, the Fermi energies of the interacting
and non-interacting systems coincide, i.e. $\omega_F=\omega_F^0$.
For fermions interacting through a hard-core potential of range $a$ in
$d=3$, Galitskii (1958) obtains $\omega_F = \omega_F^0 \big(1+ 4 k_F
a/[3\pi] + 4 (11-2\ln 2) (k_F a)^2/[15\pi^2]\big)$. Further, according
to the Seitz theorem (Seitz 1940, pp.~343 and 344; Mahan 1981, Ch.~5)
one has $\omega_F = E_0(\rho_0) + \rho_0 {\rm d} E_0(\rho_0)/{\rm
d}\rho_0$, where $E_0(\rho_0)$ stands for the ground-state total energy
per particle and $\rho_0$ the homogeneous ground-state number density.
Using the Gell-Mann and Brueckner (1957) expression for $E_0(\rho_0)$,
valid for high densities of fermions interacting through the long-range
Coulomb potential in $d=3$, or expressions derived from quantum Monte-Carlo
calculations for a wide range of densities, one can deduce that $\omega_F
\not=\omega_F^0$. }
Exception concerns the case of non-interacting fermions for which
$\Sigma(k,\omega)\equiv 0$.

From the above-presented asymptotic expression for the self-energy
of FLs, making use of Eq.~(\ref{e1}) and $G_0^{-1}(\omega) = \omega
+ \Sigma(0)$, one readily obtains
\footnote{\label{f3}
\small
Neglect of $\Sigma(\omega)$ in the denominator of Eq.~(\ref{e1}),
as done by HKM (1994) and HKMS (1994), implies the following
relation, to be compared with that in Eq.~(\ref{e2}):
$\delta G(\omega) \sim \Sigma(0)/\omega^2 + \beta_0/\omega
- i\alpha_0\, {\rm sgn}(\omega)\,\phi(\omega)$, $\omega\to 0$.}
\begin{equation}
\label{e2}
\delta G(\omega) \sim \frac{1/(1-\beta_0)}{\omega}
- i\, \frac{\alpha_0}{(1-\beta_0)^2}\,
{\rm sgn}(\omega)\,\phi(\omega) 
-\frac{1}{\Sigma(0)},\;\;\; \omega\to 0,
\end{equation}
where $\phi(\omega) \equiv -\ln\vert\omega\vert$ in $d=2$ and $\phi(\omega) 
\equiv 1$ in $d=3$. In Eq.~(\ref{e2}), $\beta_0 = 1 - 1/Z_0$ where $Z_0
\in (0,1)$ is the amount of jump discontinuity in the momentum-distribution 
function ${\sf n}(k)$ of the system at $k=0$ (at places in the following
we denote $Z_0$ by $Z_{{\Bf k}_F}$). 

% III.
\section{The Haldane bosonisation technique and its application to a
two-dimensional model of interacting fermions}

Starting from a many-body Hamiltonian for fermions interacting through 
a two-body isotropic potential, integrating out the high-energy 
degrees of freedom of the system by means of the renormalisation-group 
technique in the momentum space, Houghton and Marston (1993) have arrived 
at an effective Hamiltonian which up to regular terms --- i.e. those which 
are {\sl irrelevant} from the point of view of the renormalisation-group 
approach --- can be expressed in terms of current operators (see 
Eq.~(\ref{e21}) below and text following it). Employing Haldane's (1992) 
scheme in dealing with low-energy degrees of freedom of interacting 
fermions (i.e. making use of Haldane's `patching' scheme of a narrow band 
encompassing the FS), Houghton and Marston (1993), similar to Haldane 
(1992), establish (see Section V however) that up to some error terms 
--- which would be made small by an appropriate choice of the `patching' 
parameters --- the mentioned current operators form a Kac-Moody algebra 
(see Brink and Hanneaux 1988, Goddard and Olive 1986). This renders the 
reduced problem, as described by the indicated effective Hamiltonian, 
exactly solvable through the process of bosonisation which amounts to 
introducing a set of bosonic operators in terms of which the current 
operators can be expressed. 
\footnote{\label{f4}
\small
For completeness, we mention that the total bosonised Hamiltonian 
separates into two contributions, one involving the symmetric (or {\sl 
charge}) and the other anti-symmetric (or {\sl spin}) combination of 
the bosonic operators associated with two different spin states. In the 
treatment that we consider in the present work, the {\sl spin} part of the 
Hamiltonian is neglected. This amounts to the assumption that excitations 
in the spin sector are of higher energy than those in the charge sector.}

Although through application of the bosonisation technique one would in 
principle exactly solve the many-body problem in the low-energy region 
of the excitation spectrum (see Section V however), technical 
difficulties hinder exact determination of, e.g., $\Sigma(k,\omega)$, for 
small $\vert\omega\vert$, from the exact solution. This is owing to the 
fact that the single-particle Green function pertaining to the effective 
Hamiltonian, which we for clarity denote by $G_{\rm f}$ (subscript 
${\rm f}$ making explicit that this Green function pertains to the actual 
{\sl fermion} system) in the ${\Bf k},\omega$ space is to be obtained 
from the Fourier transform with respect to the reciprocal variables 
${\Bf x},t$ (space-time) of an expression in which the 
${\Bf x},t$-representation of the exact Green function of the bosonic 
problem, which we denote by $G_{\rm b}$, is exponentiated. It should be 
noted that in the context of Haldane's (1992) bosonisation scheme, one 
obtains $G_{\rm b}({\Bf k},\omega;S)$ with $S$ indicating the FS `patch' 
to whose corresponding `squat box' (Houghton and Marston 1993) ${\Bf k}$ 
pertains. In $d$ dimensions, $S$ is a $(d-1)$-dimensional 
flat surface of linear size $\Lambda$, thus covering an area of extent 
$\Lambda^{d-1}$. The `squat box' associated with $S$ is defined as the 
region around ${\Bf k}_F(S)$, the Fermi wave-vector at the centre of $S$, 
which consists of {\sl all} points of `patch' $S$ translated by $\gamma\,
\mbox{\boldmath{$n$}}_S$ with $\gamma \in [-\lambda/2,\lambda/2]$, 
$\lambda\ll\Lambda$; here $\mbox{\boldmath{$n$}}_S$ denotes the outward 
unit vector normal to $S$. For definiteness, let ${\cal F}_{{\Bf k}, 
\omega}[f] {=:} g({\Bf k},\omega)$ denote the Fourier transform of 
$f({\Bf x},t)$ to the ${\Bf k},\omega$ domain and ${\cal F}^{-1}_{{\Bf 
x},t}[g] \equiv f({\Bf x},t)$ its inverse. Let further 
\footnote{\label{f5}
\small
One can equivalently write ${\tilde {\cal F}}^{-1}_{{\Bf x},t}[g] {:=} 
{\cal F}^{-1}_{{\Bf x},t}[g]-{\cal F}^{-1}_{{\Bf 0},0}[g]$. We point
out that in the case at hand, in particular for $d=2$, it can be shown that 
$\delta G_{\rm b}({\Bf x},t;S)$ is continuous at both ${\Bf x}={\Bf 0}$ 
and $t=0$ so that ${\cal F}^{-1}_{{\Bf 0},0}[\delta G_{\rm b}]$ is
{\sl not} ambiguous.}
${\tilde {\cal F}}^{-1}_{{\Bf x},t}[g] {:=} {\cal F}^{-1}_{{\Bf x},t}[g] - 
f({\Bf 0},0)$. For $d=2$, $G_{\rm f}({\Bf k},\omega;S)$ is to be obtained 
from (HKM 1994, HKMS 1994)
\begin{equation}
\label{e3}
G_{\rm f}({\Bf k},\omega;S)
= {\cal F}_{{\Bf k},\omega}
\Big[\frac{\Lambda}{(2\pi)^2}\,
\frac{\exp\big(i {\Bf k}_F(S)\cdot 
{\Bf x}\big)}{ {\Bf x}\cdot\mbox{\boldmath{$n$}}_S - t + 
i\,\eta\,{\rm sgn}(t)}\,\exp\big(\frac{2\pi i}{\Omega^2}
{\tilde {\cal F}}^{-1}_{{\Bf x},t}[\delta G_{\rm b}]\big)\Big], \;\;\;
\vert {\Bf x}\cdot \mbox{\boldmath{$\tau$}}_S\vert \ll 1/\Lambda,
\end{equation}
where $\Omega {:=} \Lambda (L/[2\pi])^2$ with $L$ the (macroscopic) linear 
size of the square in which the system is confined and
$\mbox{\boldmath{$\tau$}}_S$ is the unit vector normal to
$\mbox{\boldmath{$n$}}_S$; $\delta G_{\rm b}$ denotes deviation of the 
full bosonic Green function from the free one. Note in passing that since 
each FS `patch' is flat, ${\Bf k}_F(S) = k_F(S)\, 
\mbox{\boldmath{$n$}}_S$. The expression in Eq.~(\ref{e3}) makes 
explicit the complexity of calculation of $G_{\rm f}({\Bf k},\omega;S)$ 
from $G_{\rm b}({\Bf k},\omega;S)$. 

For our following considerations it is useful to define
\begin{equation}
\label{e3a}
\Xi {:=} \exp\big(\frac{-2\pi i}{\Omega^2} 
{\cal F}^{-1}_{{\Bf x}={\Bf 0},t=0}[\delta G_{\rm b}]\big).
\end{equation}
It is important to realise that $\Xi\not=0$, for $\Xi=0$ implies 
$G_{\rm f}({\Bf k},\omega;S)\equiv 0$ which is meaningless. Following 
the above definition for $\Xi$, for $G_{0;{\rm f}}({\Bf k},\omega;S)$ 
and $\delta G_{\rm f}({\Bf k},\omega;S)$ in $G_{\rm f}({\Bf k},
\omega;S) \equiv G_{0;{\rm f}}({\Bf k},\omega;S) +\delta G_{\rm f}
({\Bf k},\omega;S)$ (compare with $G_0(k,\omega)$ and $\delta G(k,
\omega)$ introduced above) we have
\footnote{\label{f6}
\small
We point out that our definition of $\delta G_{\rm f}({\Bf k},\omega;S)$ 
differs from that in the works by HKM (1994) and HKMS (1994). These authors 
employ the first-order expansion $\exp\big(2\pi i\Omega^{-2} {\tilde 
{\cal F}}^{-1}_{{\Bf x},t}[\delta G_{\rm b}]\big) \approx 1 + 2\pi 
i\Omega^{-2}{\tilde {\cal F}}^{-1}_{{\Bf x},t}[\delta G_{\rm b}]$ and 
tacitly dispose of the contribution ${\cal F}^{-1}_{{\Bf 0},0}[\delta 
G_{\rm b}]$ associated with the latter ${\tilde {\cal F}}^{-1}_{{\Bf x},
t}[\delta G_{\rm b}]$; thus in these author's work $\delta G_{\rm f}
({\Bf k},\omega;S) \approx {\cal F}_{{\Bf k},\omega}\big[(2\pi)^{-2} 
\Lambda\, \big\{\exp\big(i {\Bf k}_F(S)\cdot {\Bf x}\big)/[{\Bf x}\cdot
\mbox{\boldmath{$n$}}_S - t + i\,\eta\,{\rm sgn}(t)]\big\}\, 2\pi 
i\Omega^{-2} {\cal F}^{-1}_{{\Bf x},t}[\delta G_{\rm b}]\big]$, whereas 
using the same linear expansion we have $\delta G_{\rm f}({\Bf k},
\omega;S)\approx\big(\Xi -1)\, G_{0;{\rm f}}({\Bf k},\omega;S) + \Xi 
{\cal F}_{{\Bf k},\omega}\big[(2\pi)^{-2} \Lambda\, \big\{\exp\big(i 
{\Bf k}_F(S)\cdot {\Bf x}\big)/[{\Bf x}\cdot\mbox{\boldmath{$n$}}_S - 
t + i\,\eta\,{\rm sgn}(t)]\big\}\, 2\pi i\Omega^{-2} {\cal F}^{-1}_{{\Bf 
x},t} [\delta G_{\rm b}]\big]$. Therefore, for properties which are 
associated with interaction, our (first-order) results can be directly 
compared with those by HKM (1994) and HKMS (1994) through dividing our 
results by $\Xi$, which, as we have indicated, is non-vanishing. }
\begin{equation}
\label{e3b}
G_{0;{\rm f}}({\Bf k},\omega;S) {:=}
{\cal F}_{{\Bf k},\omega}\Big[ \frac{\Lambda}{(2\pi)^2}
\,\frac{\exp\big(i {\Bf k}_F(S)\cdot {\Bf x}\big)}{ {\Bf x}\cdot 
\mbox{\boldmath{$n$}}_S -t + i\,\eta\,{\rm sgn}(t)}\Big],
\;\;\; \vert {\Bf x}\cdot \mbox{\boldmath{$\tau$}}\vert \ll 1/\Lambda,
\end{equation}
\begin{eqnarray}
\label{e3c}
& &\delta G_{\rm f}({\Bf k},\omega;S) {:=}
(\Xi -1)\, G_{0;{\rm f}}({\Bf k},\omega;S)\nonumber\\
& &\;\;\;\; 
+ \Xi\, {\cal F}_{{\Bf k},\omega}\Big[ \frac{\Lambda}{(2\pi)^2}
\,\frac{\exp\big(i {\Bf k}_F(S)\cdot {\Bf x}\big)}{ {\Bf x}\cdot 
\mbox{\boldmath{$n$}}_S - t + i\,\eta\,{\rm sgn}(t)}
\big\{\exp\big(\frac{2\pi i}{\Omega^2} {\cal F}^{-1}_{{\Bf x},t}
[\delta G_{\rm b}]\big) - 1\big\}\Big],
\;\;\;
\vert {\Bf x}\cdot \mbox{\boldmath{$\tau$}}\vert \ll 1/\Lambda.
\end{eqnarray}
It is readily verified that for ${\Bf k} - {\Bf k}_F(S)$ parallel to 
$\mbox{\boldmath{$n$}}_S$ (i.e. for ${\Bf k} - {\Bf k}_F(S) = k_{\parallel}\,
\mbox{\boldmath{$n$}}_S$), with the range of the ${\Bf x}$-integration 
along $\mbox{\boldmath{$\tau$}}_S$ restricted to interval 
$(-\pi/\Lambda,\pi/\Lambda)$, one has
\begin{equation}
\label{e3d}
G_{0;{\rm f}}({\Bf k},\omega;S) = \frac{1}{\omega - k_{\parallel}
+ i\,\eta\,{\rm sgn}(\omega)},\;\;\; \eta\downarrow 0,
\end{equation}
so that for $k_{\parallel}=0$ the first term on the right-hand side (RHS)
of Eq.~(\ref{e3c}) is seen to behave like $(\Xi-1)/\omega$ as $\omega\to 
0$. Thus {\sl if} for $\omega\to 0$ the leading asymptotic contribution 
due to the second term on the RHS of Eq.~(\ref{e3c}) is sub-dominant with 
regard to $1/\omega$, a comparison with Eq.~(\ref{e2}) reveals that for 
conventional FLs must hold
\footnote{\label{f7}
\small
Here we are relying on a theorem which asserts uniqueness of coefficients 
in an asymptotic series expansion of a function with respect to a given 
asymptotic sequence (see Lauwerier 1977). }
$\Xi=1/(1-\beta_0) + 1 \equiv Z_{{\Bf k}_F} +1$. For unconventional
FLs, with $\Sigma(0)=0$, on the other hand, $\Xi =\beta_0/(1-\beta_0)+1
\equiv Z_{{\Bf k}_F}$ (see Eq.~(\ref{e7}) below). It is important to 
realise that these relationships do {\sl not} apply to NFLs, for our 
inference has been based upon Eq.~(\ref{e2}) above and Eq.~(\ref{e7}) 
below which are specific to FLs. In this connection we mention that for 
LLs the leading contribution to ${\rm Re}\delta G(\omega)$, which at the 
same time is the leading contribution to $\delta G(\omega)$, is {\sl 
exactly} equal to $-1/\omega$, and for marginal FLs (Varma, {\sl et al.}, 
1989, 1990, Littlewood and Varma 1991), depending on whether $\Sigma(0)=0$ 
or $\Sigma(0)\not=0$, one has (exactly) $-1/\omega$ and 
$[-\Sigma(0)/\beta_0]/(\omega\ln\vert\omega\vert)$, $\omega\to 0$, 
respectively; here $\beta_0$ is the coefficient of $\omega\ln\vert\omega
\vert$ in the asymptotic expansion of $\Sigma(\omega)$ for $\omega\to 0$. 
Since $\Xi\not=0$, in cases where the leading asymptotic contribution to 
$\delta G(\omega)$ is equal to $-1/\omega$, for $\omega\to 0$, the 
leading asymptotic contribution of the second term on the RHS of 
Eq.~(\ref{e3c}) must be $-\Xi/\omega$, for $\omega\to 0$, thus cancelling 
the $\Xi/\omega$ contribution due to the first term. As we shall discuss 
below, within the approximation where $\exp\big(2\pi i\Omega^{-2} 
{\cal F}^{-1}_{{\Bf x},t}[\delta G_{\rm b}]\big) \approx 1 + 2\pi 
i\Omega^{-2} {\cal F}^{-1}_{{\Bf x},t}[\delta G_{\rm b}]$, our explicit 
calculations establish that the second term on the RHS of Eq.~(\ref{e3c}) 
scales like $\ln^2\vert\omega\vert$ for $\omega\to 0$ so that, provided 
the latter approximation be justified, we are led to the conclusion that
under {\sl no} circumstance (i.e. independent of the nature of the
particle-particle interaction) the system under consideration can be either 
a LL or a marginal FL. It is interesting to note that $\ln^2\vert\omega
\vert$ does not coincide with any of the known next-to-leading asymptotic 
results (presented above) corresponding to $\delta G(\omega)$. This aspect
may be viewed as indicating the inadequacy of the first-order approximation 
$\exp\big(2\pi i\Omega^{-2}{\cal F}^{-1}_{{\Bf x},t}[\delta G_{\rm b}]\big) 
\approx 1 + 2\pi i\Omega^{-2} {\cal F}^{-1}_{{\Bf x},t}[\delta G_{\rm b}]$ 
as well as that of the Haldane (1992) bosonisation scheme in $d > 1$. 

% III.a
\subsection{Simplified and approximate approaches}

Due to the mentioned complexity of calculation of $G_{\rm f}({\Bf k},
\omega;S)$, HKM (1994) and HKMS (1994) employ two expansions of which 
one in certain limits is in principle rigorous.
\footnote{\label{f8}
\small
Our cautious remark reflects our experience in dealing with the problem 
at hand where the most innocuous approximations turn out to change even 
the qualitative aspects of the results (Farid 1999b).}
It is therefore the second of these expansions that will concern us most 
in our following considerations. As for the first expansion, in principle
rigorous in certain limits, HKM (1994) and HKMS (1994) proceed from the 
exact expression for the proper self-energy of the bosonic problem, namely 
\begin{equation}
\label{e4}
\Sigma_{\rm b}({\Bf q},\omega;S) 
=\frac{2\Lambda^{d-1} \mbox{\boldmath{$n$}}_S\cdot {\Bf q}}
{(2\pi)^d}\, W({\Bf q},\omega),\;\;\;\;\;\;
W({\Bf q},\omega) {:=}
\frac{v({\Bf q})}{1 + v({\Bf q}) \chi_0({\Bf q},
\omega)},
\end{equation}
where $v({\Bf q})$ stands for the Fourier transform of the two-body 
interaction potential and $\chi_0({\Bf q},\omega) \equiv {\bar\chi}_0
(\omega/\|{\Bf q}\|)$, with ${\bar\chi}_0(x)$ the Lindhard function. 
The Lindhard function employed here and by HKM (1994) and HKMS (1994) 
is evaluated under the assumption that $\Lambda\downarrow 0$, taking 
into account the requirement $\lambda \ll \Lambda$; in this limit, a 
{\sl sum} over FS `patches' can be replaced by an {\sl integral} (see 
Eq.~(\ref{e17a}) below). HKM (1994) consider the case of a short-range 
potential, replacing $v({\Bf q})$ by $f_0$ which they assume to be small, 
thus justifying their use of $f_0/[1 + f_0{\bar\chi}_0(x)] \approx f_0 
- f_0^2 {\bar\chi}_0(x)$; $N(0) f_0$, with $N(0) \equiv k_F/[2\pi]$ the 
density of states at the Fermi energy, may be considered as the $\ell=0$ 
component of the symmetric or charge part of the Landau parameter 
$F_{\ell}^s$. HKMS (1994), on the other hand, consider the cases of the 
long-range Coulomb potential and super long-range potentials $v({\Bf q}) 
= g/\|{\Bf q}\|^{\gamma}$ with $g$ the coupling constant of interaction; 
in $d=2$, the former corresponds to $\gamma=1$ and the latter to $\gamma 
> 1$ and in the particular case of a logarithmic interaction to $\gamma=2$. 
In these cases, corresponding to $\gamma=1,2$, taking into account the 
diverging behaviour of $v({\Bf q})$ as ${\Bf q}\to {\Bf 0}$, these 
authors employ
\footnote{\label{f9}
\small
The limit $v({\Bf q})\to\infty$ must be taken {\sl after} evaluation of 
a pertinent integral over $\omega'$, for otherwise, in consequence of 
$1/{\bar\chi}_0(x)\sim - 2 x^2/N(0)$, for $\vert x\vert\to\infty$, the 
expression for $\delta G(k_{\parallel},\omega)$ will involve a contribution 
proportional to $\int_{-\lambda_{\parallel}/2}^{\lambda_{\parallel}/2} 
{\rm d}q_{\parallel}\int_{-\lambda_{\perp}/2}^{\lambda_{\perp}/2} 
{\rm d}q_{\perp}\,[1/N(0)]/\|{\Bf q}\|^2$ (due to semi-circle contours 
at infinity in the complex $\omega'$-plane) which is spurious. In the 
work by HKMS (1994) this contribution has been inadvertently (but 
correctly) neglected. }
$v({\Bf q})/[1+v({\Bf q}){\bar\chi}_0(x)] \approx 1/{\bar\chi}_0(x)$. 

The second expansion employed by HKM (1994) and HKMS (1994) consists 
in that of the second exponential function on the RHS of Eq.~(\ref{e3}) 
which these authors carry out to linear order (see footnote \ref{f6}); 
this liner expansion has been employed by the authors subsequent to 
their use of the pertinent expansion which we have indicated in the 
previous paragraph. For simplicity, but without loss of generality, 
here we only explicitly deal with the result corresponding to the case 
in which $v({\Bf q})$ is replaced by $f_0$. For the leading contribution 
(as specified by the lowest relevant power of $f_0$) to the {\sl imaginary} 
part of $\delta G_{\rm f}(k_{\parallel},\omega;S)$, HKM (1994) obtain (for 
$d=2$; below we suppress $S$ in order to conform with the notation by HKM)
\footnote{\label{f10}
\small
With reference to our remark in footnote \ref{f6}, the RHS of this 
expression would have been multiplied by $\Xi$ if our definition
for $\delta G(\omega)$ had been employed.}
\begin{equation}
\label{e5}
{\rm Im}\delta G_{\rm f}(k_{\parallel},\omega) = 
G_0(k_{\parallel},\omega) G_0(k_{\parallel},\omega) 
{\cal G}(k_{\parallel},\omega),
\end{equation}
\begin{eqnarray}
\label{e6}
{\cal G}(k_{\parallel},\omega) &{:=}&
\frac{1}{2}\,\frac{f_0^2 N(0)}{(2\pi)^2}
\,{\rm sgn}(\omega)\,
\Big\{ [\omega^2 + (\omega-k_{\parallel})^2/4]
\ln\frac{\vert \omega-k_{\parallel}\vert}{\lambda_{\perp}}
\nonumber\\
& &\;\;\;\;\;\;\;\;\;\;\;\;\;
+[\omega^2 - (\omega-k_{\parallel})^2/4]
\ln\frac{\vert \omega+k_{\parallel}\vert}{\lambda_{\perp}}
-\frac{1}{2}\omega (2\omega-k_{\parallel})\Big\},
\end{eqnarray}
where $\lambda_{\perp}$ denotes a cut-off parameter restricting momentum 
integrations in the direction perpendicular to $\mbox{\boldmath{$n$}}_S$
to the interval $[-\lambda_{\perp}/2,\lambda_{\perp}/2]$. By neglecting 
$\Sigma(k,\omega)$ in the denominator of the expression on the RHS of 
Eq.~(\ref{e1}), HKM (1994) identify ${\cal G}(k_{\parallel},\omega)$ with 
${\rm Im}\Sigma_{\rm f}(k_{\parallel},\omega)$ to second order in $f_0$. 

% III.b
\subsection{Some noteworthy observations}

In the work by HKM (1994) as well as that by 
HKMS (1994), the authors apparently identify the zero of energy $\omega$ 
with the Fermi energy of the {\sl non}-interacting system, i.e. 
$\omega_F^0 = 0$. In this sense the Green function $G_0$ on the RHS of 
Eq.~(\ref{e5}) is in principle to be distinguished from $G_0$ as employed 
in our above considerations; HKM mention namely: ``The location of the 
[quasi-particle] pole has been shifted from its bare value to $\omega = 
v_F'\,k_{\parallel}$ due to renormalization of the Fermi velocity ...''. 
The authors thus fix the Fermi energy of the interacting system by 
replacing $k_{\parallel}$ in Eq.~(\ref{e6}) by $\omega/v_F'$, where 
$v_F' = 1 + F_0^s (1-F_0^s)\Lambda/[2\pi k_F]$. Subsequently, making use 
of $1/v_F' \approx 1 - F_0^s \Lambda/[2\pi k_F]$, HKM (1994) obtain the 
imaginary part of the on-the-mass-shell self-energy, denoted by 
${\rm Im}\Sigma_{\rm f}^{(2)}(\omega)\vert_{\rm pole}$, which is of the 
form ${\rm sgn}(\omega)\{ A\omega^2\ln\vert\omega\vert + B\omega^2\}$; 
for the explicit forms of $A$ and $B$ see Eq.~(63) in (HKM 1994). As we 
shall now demonstrate, the finding with regard to the quasi-particle pole 
having been shifted is based on an incorrect observation. To this end we 
first point out that ${\rm Im}\Sigma_{\rm f}^{(2)}(k_{\parallel},\omega)$ 
stripped off of ${\rm sgn}(\omega)$ is zero at $\omega=0$, irrespective 
of the value chosen for $k_{\parallel}$, and moreover is negative 
elsewhere. The change of sign by ${\rm Im}\Sigma_{\rm f}^{(2)}
(k_{\parallel},\omega)$ is therefore entirely due to ${\rm sgn}(\omega)$. 
Stability of the system implies (and this is at the same time a corollary 
to a theorem due to Luttinger (1960) --- see also Luttinger and Ward
(1960)) that $\omega=0$ {\sl must} be the Fermi energy of the interacting 
system. From this it follows that ${\rm Re}\Sigma_{\rm f}(0,0) =0$, 
which in turn implies that the self-energy as calculated by HKM (1994)
(and by HKMS (1994)) {\sl cannot} correspond to a conventional FL (see 
text following Eq.~(\ref{e1}) above). In fact, as we shall demonstrate 
in this work, within the Haldane (1992) bosonisation approach, under 
{\sl no} circumstance the system under consideration can be a conventional 
FL; neither can it be a LL, {\sl even} for super-long-range interactions
(see Section IV.B).

The property ${\rm Re}\Sigma_{\rm f}(0,0) =0 \iff \Sigma(0)=0$
enables us to identify $G_0(k_{\parallel},\omega)$ in Eq.~(\ref{e5}) 
with that employed by us at the outset of this work. Further, in
consequence of $\Sigma(0)=0$, the appropriate expression for
$\delta G(\omega)$ is
\footnote{\label{f11}
\small
Neglect of $\Sigma(\omega)$ in the denominator of Eq.~(\ref{e1}), as done 
by HKM (1994) and HKMS (1994), implies the following relation, to be 
compared with that in Eq.~(\ref{e7}): $\delta G(\omega) \sim \beta_0/\omega 
- i\alpha_0\, {\rm sgn}(\omega)\,\phi(\omega)$, $\omega\to 0$.}
\begin{equation}
\label{e7}
\delta G(\omega) \sim \frac{\beta_0/(1-\beta_0)}{\omega}
- i\, \frac{\alpha_0 (1+\beta_0)}{1-\beta_0}\, {\rm sgn}(\omega)
\,\phi(\omega),
\end{equation}
rather than that in Eq.~(\ref{e2}). Although the leading contributions 
in Eqs.~(\ref{e2}) and (\ref{e7}) both scale like $1/\omega$, it must 
be noted that the coefficient of $1/\omega$ in Eq.~(\ref{e2}), namely 
$1/(1-\beta_0)$, is positive while that in Eq.~(\ref{e7}), i.e. 
$\beta_0/(1-\beta_0)$, is negative. In view of remarks following
Eq.~(\ref{e3d}) above, we conclude that {\sl if} the system under 
consideration is a FL, even though an unconventional one, then we have 
$\Xi=Z_{{\Bf k}_F}$.

We have examined ${\rm Re}\delta G(\omega)$ --- calculated independently,
employing the same approximate treatment of Eq.~(\ref{e3}) as employed by 
HKM (1994) and HKMS (1994) in their calculation of ${\rm Im}\delta 
G(\omega)$ --- pertaining to the model under consideration in considerable 
detail (results to be published shortly --- Farid 1999b) and found that
the second term on the RHS of Eq.~(\ref{e3c}) has {\sl no} contribution to 
${\rm Re}\delta G(\omega)$, for $\omega\to 0$, scaling like $1/\omega$ 
(see Eqs.~(\ref{e2}) and (\ref{e7}) above). Our rigorous calculations show 
that contribution of this term to ${\rm Re}\delta G(\omega)$ scales like 
$\ln^2\vert\omega\vert$ for $\omega\to 0$; this result concerns the case 
where $v({\Bf q})\to\infty$ --- only for this case have we been able to 
perform our entire calculations fully analytically ---, however 
numerically-obtained results corresponding to a general $v({\Bf q})$, 
including super long-range interactions, reveal a similar behaviour. With 
reference to our remarks following Eq.~(\ref{e3d}) above, we can therefore 
conclude that provided the first-order expansion of the exponential
function within braces on the RHS Eq.~(\ref{e3c}) suffice (which does not 
seem to be the case, at least not for situations where the particle-particle 
interaction is super-long-range --- see Section IV), the system under 
consideration must be an unconventional FL with $\Sigma(0)=0$ and 
$Z_{{\Bf k}_F}=\Xi$.

Two comments are in order here. First, if $\Sigma(0)\not=0$ {\sl and} 
$\Sigma(\omega)$ in the denominator of Eq.~(\ref{e1}) were to be 
neglected (both of these conditions are implicit in the works by HKM 
(1994) and HKMS (1994)), then the leading asymptotic contribution to 
$\delta G(\omega)$ would diverge like $1/\omega^2$ for $\omega\to 0$ 
(see footnote \ref{f3}). This would imply, since $\delta G(\omega)$ would 
not be integrable (in the Riemann sense) in a neighbourhood of $\omega=0$, 
that {\sl no} spectral representation for $\delta G(\omega)$ would exist 
(Farid 1999b), contradicting an obvious fact. Second, if $\Sigma(0)=0$
{\sl and} $\Sigma(\omega)$ in the denominator of Eq.~(\ref{e1}) were to 
be neglected, then Eq.~(\ref{e1}) would imply $\Sigma(\omega)\sim\omega^2
\delta G(\omega)$. From the spectral representation for $\delta G(\omega)$, 
on the other hand, it follows that to ${\rm Im}\delta G(\omega) \sim A\,
{\rm sgn}(\omega)\,\ln\vert\omega\vert$ corresponds ${\rm Re}\delta 
G(\omega) \sim C$, for $\omega \to 0$, with $C$ a constant (which may or 
may not be vanishing). Thus $\Sigma(\omega)\sim\omega^2\delta G(\omega)$ 
would imply not only ${\rm Im}\Sigma(\omega) \sim A\, {\rm sgn}(\omega)\,
\omega^2\,\ln\vert\omega\vert$, but also ${\rm Re}\Sigma(\omega)\sim C\,
\omega^2$, for $\omega\to 0$. Now whereas ${\rm Im}\Sigma(\omega) \sim A\, 
{\rm sgn}(\omega)\,\omega^2\, \ln\vert\omega\vert$ would lead one to 
consider the metallic state at issue to be a FL, this would be at odds 
with ${\rm Re}\Sigma(\omega) \sim C\,\omega^2$, for, as we have mentioned 
in the paragraph preceding Eq.~(\ref{e1}) above, in the asymptotic 
series of ${\rm Re}\Sigma(\omega)$, for $\omega\to 0$, pertaining to
{\sl interacting} FLs (both conventional and unconventional FLs), there 
{\sl must} occur a term linear in 
\footnote{\label{f12}
\small
According to Eq.~(\ref{e7}), which holds for FLs with $\Sigma(0)=0$,
$\omega^2 \delta G(\omega) \sim [\beta_0/(1-\beta_0)] \omega$. Since
we have shown $\Xi = Z_{{\Bf k}_F} \equiv 1/(1-\beta_0)$ under {\sl all}
circumstances as regards nature of $v({\Bf q})$, vanishing of this linear 
term would amount to $Z_{{\Bf k}_F} = 1$, or $\Xi = 1$ which through 
Eq.~(\ref{e3a}) would imply ${\cal F}^{-1}_{{\Bf x}={\Bf 0},t=0}[\delta 
G_{\rm b}] = 0$. Evidently, this can occur {\sl only} if $v({\Bf q})
\equiv 0$. }
$\omega$. These two comments make evident that in deducing $\Sigma(\omega)$ 
from $\delta G(\omega)$, in particular in the cases where one has to do 
with FLs, the self-energy in the denominator of Eq.~(\ref{e1}) must not 
be neglected. 

As our last observation in this Section, we mention that since ${\rm Im}
\delta G(\omega)\sim [-f_0^2 N(0)/(2\pi)^2]\, {\rm sgn}(\omega) 
\ln\vert\omega\vert$ has been derived from the first-order expansion of 
the exponential function on the RHS of Eq.~(\ref{e3}) (see footnote 
\ref{f6}), taking into account that apparently $\Sigma(0)=0$ --- a 
property which, though not exclusively, is specific to LLs ---, it is 
tempting to assume that higher-order contributions that correspond to 
the higher-order terms in the expansion of the mentioned exponential 
function involve appropriate powers of this first-order result, 
\footnote{\label{f13}
\small
A cursory analysis suggests this to be the case. However, close inspection 
of this procedure reveals that a rigorous justification for exponentiating 
the above first-order result is far from trivial and, as we shall indirectly
demonstrate in this work, incorrect. We note in passing that it {\sl 
appears} to us that Bares and Wen (1993) employ a similar procedure in 
dealing with the low-energy behaviour of the single-particle Green function 
(see the paragraph following Eq.~(42) in Bares and Wen 1993). If this 
indeed is the case, then in our opinion the conclusion arrived at by Bares 
and Wen (1993) may be of questionable nature. }
thus allowing one to exponentiate the indicated first-order result and 
obtain a power-law behaviour for the self-energy. Explicitly, by pursuing 
this temptation, one obtains $\Sigma(\omega)\sim \beta_0 [\cot(\pi\gamma_0) 
+ i]\, {\rm sgn}(\omega) \vert\omega\vert^{1-2\gamma_0}$, for $\omega
\to 0$, where 
\footnote{\label{f14}
\small
Using our definition for $\delta G_{\rm f}$, in the following two
expressions $f_0^2$ need be replaced by $\Xi f_0^2$.}
\begin{equation}
\label{e8}
\gamma_0= \frac{1}{2}\big[1- \frac{1}{(2\pi)^2} f_0^2 N(0)\big],
\;\;\;\;\;\;\;
\beta_0 = \frac{-1}{2 (2\pi)^2}\, f_0^2 N(0)\,
\left[\cot(\pi\gamma_0) + 1\right].
\end{equation}
The above asymptotic expression for $\Sigma(\omega)$ is that of LLs, with 
$\alpha\equiv 2\gamma_0$ the anomalous dimension (see, e.g., Farid 1999a). 
As it will become evident, this result is not tenable. However, the prospect 
that this result offers, that a LL metallic state would be feasible in 
$d=2$, has been a further impetus for us to investigate the problem at 
hand in more detail, whence our exact treatment in Section IV.

% IV
\section{The long-distance behaviour of the single-particle Green
function; an exact treatment}

Exact numerical evaluation of $\Sigma(k,\omega)$ within the Haldane (1992) 
bosonisation framework is in principle feasible for $d=2$ but involves a 
considerable amount of investment in computational efforts as well as 
computational resources. For our purpose, which at present in the main 
consists of establishing whether the metallic state of the model under 
consideration is a FL or otherwise, it is sufficient to determine the 
leading term in the large-$\|{\Bf x}\|$ asymptotic expansion of the 
single-particle Green function in the ${\Bf x},t$-representation for $t$ 
infinitesimally negative, i.e. $G({\Bf x},t=0^-)$, which can be achieved 
fully analytically. 

\subsection{General considerations}

As can be seen from Eq.~(\ref{e3}), for determination of the leading 
large-$\|{\Bf x}\|$ term in the asymptotic series of $G({\Bf x},t=0^-)$
it is sufficient to calculate ${\rm Re}\big(2\pi i\Omega^{-2}{\cal 
F}^{-1}_{{\Bf x}, t=0^-}[\delta G_{\rm b}]\big)$ for large $\|{\Bf x}\|$. 
Before proceeding with this, we present some results which will set the 
stage for what follows. First, in $d=1$ we have (see, e.g., Farid 1999b 
and Eq.~(\ref{e11e}) below)
\begin{equation}
\label{e9}
-i G(x,t=0^-) \sim \frac{Z_{k_F}}{\pi}\, \frac{\sin(k_F x)}{x}\,
\;\;\;\mbox{\rm for}\;\;
\vert x\vert\to \infty\;\;\;\;\;\mbox{\rm (FLs, $d=1$)},
\end{equation}
\begin{equation}
\label{e10}
-i G(x,t=0^-) \sim \big[-2i\alpha C \int_0^{\Delta}
\frac{{\rm d} k}{2\pi}\; k^{\alpha-1}\,\cos(k)\big]\,
\frac{{\rm sgn}(x)\,\exp(i k_F x)}{\vert x\vert^{1+\alpha}},  
\;\mbox{\rm for}\;
\vert x\vert\to \infty\;\mbox{\rm (LLs, $d=1$)}.
\end{equation}
We note that the asymptotic result pertaining to the LL in $d=1$ can be 
directly inferred from the explicit expression for $G(x,t)$ pertaining 
to the one-dimensional Luttinger model (Luttinger 1963, Mattis and Lieb 
1965) which is available (see, e.g., Voit 1994). For the calculation of 
the result in Eq.~(\ref{e10}) we have {\sl not} made use of this exact 
result, but employed the relationship between $-i G(x,t=0^-)$ and the 
momentum distribution function ${\sf n}(k)$ (see Eq.~(\ref{e11a}) below). 
In Eq.~(\ref{e10}), $\alpha \equiv 2\gamma_0$, with $0 <\alpha < 1$, is 
the anomalous dimension (see Eq.~(\ref{e8})), $C$ a constant which 
features in the asymptotic series expansion of the momentum-distribution 
function ${\sf n}(k)$ for $k\to k_F$, {\sl viz.} ${\sf n}(k) \sim 1/2 - 
C\,{\rm sgn}(k-k_F)\,\vert k - k_F\vert^{\alpha}$. Further, $\Delta > 0$ 
denotes a cut-off parameter whose value is on the order of the width of 
the derivative with respect to $k$ of ${\sf n}(k)$ in the neighbourhood 
of $k=k_F$; the precise value of $\Delta$ is of no relevance to our work 
and it can be entirely disposed of if one employs the complete description 
of ${\sf n}(k)$ rather than its leading asymptotic terms for $k\to k_F$. 
That the RHS of Eq.~(\ref{e10}), in contrast to that of Eq.~(\ref{e9}), 
is complex-valued, has its origin in the fact that the expression in 
Eq.~(\ref{e10}) concerns only the branch of the right-movers and therefore 
is determined by a single Fermi point, namely that at $k=k_F$. 

For FLs in $d=2$ one has (see, e.g., Farid 1999b)
\begin{equation}
\label{e11}
-i G({\Bf x},t=0^-) \sim
\frac{Z_{{\Bf k}_F}\,\|{\Bf k}_F\|^{1/2}}{2^{1/2} \pi^{3/2}}\,
\frac{\sin\big(\|{\Bf k}_F\|\,\|{\Bf x}\| 
-\pi/4\big)}{\|{\Bf x}\|^{3/2}},\;\;\;\mbox{\rm for}\;\;
\|{\Bf x}\|\to \infty\;\;\;\;\;\mbox{\rm (FLs, $d=2$)},
\end{equation}
while, in analogy with the expressions in Eqs.~(\ref{e9}) and (\ref{e10}), 
for LLs in $d=2$ one has a similar expression as in Eq.~(\ref{e11}), with 
the power of $\|{\Bf x}\|$, however, changed from $3/2$ to $3/2+\alpha$ 
with $0 < \alpha < 1$. We point out that the power $3/2$ of $\|{\Bf x}\|$ 
in $d=2$, both for FLs and LLs, is composed of $1$ and $1/2$, of which $1$ 
(as well as $\alpha$ in the case of LLs) is the contribution from the radial 
part of a two-dimensional integral carried out in the cylindrical coordinate 
system and $1/2$ from the angular part. 

For our further investigations it is important to understand the 
underlying reason for the qualitative difference in the long-distance 
behaviour of $-i G({\Bf x},t=0^-)$ pertaining to FLs and LLs. Here we only 
briefly consider this subject and for a detailed discussion refer the reader 
to our forthcoming publication (Farid 1999b). First, we point out that
\begin{equation}
\label{e11a}
-i G({\Bf x},t=0^-) = \int \frac{{\rm d}^d{\Bf k}}{(2\pi)^d}\;
{\rm e}^{i {\Bf k}\cdot {\Bf x}}\, {\sf n}({\Bf k}).
\end{equation}
With $E_M$ the 
total energy of the interacting $M$-particle ground state, one can show 
that $\mu_N {:=} E_N - E_{N-1}$, with $N$ the actual number of particles 
in the system, is equal to $\omega_F$ and that $\mu_N < \mu_{N+1}$, where
$\mu_{N+1} {:=} E_{N+1} - E_N$. The latter inequality holds also in the 
thermodynamic limit, even though in this limit $\mu_{N+1}-\mu_N$ is 
infinitesimally small in the case of metals. Owing to this property, we 
can introduce $\mu$, `chemical potential', satisfying $\mu_N < \mu < 
\mu_{N+1}$. Suppose $G({\Bf k},\omega)$ be {\sl unbounded} at 
$\omega_{\Bf k}$ ($\omega_{\Bf k}$ may or may not be a pole, i.e. an 
isolated singularity) and let ${\cal C}_{\Bf k}$ denote a circle of 
infinitesimally small radius in the complex energy plane ($z$-plane) 
centred around $\omega_{\Bf k}$. In view of association of 
${\sf n}({\Bf k})$ with $-i G({\Bf x},t=0^-)$, only those $\omega_{\Bf k}$ 
are relevant to ${\sf n}({\Bf k})$ which satisfy $\omega_{\Bf k} \le 
\omega_F < \mu$. One has
\begin{equation}
\label{e11b}
{\sf n}({\Bf k}) = \Theta(\mu - \omega_{\Bf k})\,\int_{{\cal C}_{\Bf k}}
\frac{{\rm d} z}{2\pi i}\; G({\Bf k},z) + {\wp}\int_{-\infty}^{\mu}
\frac{{\rm d}\omega}{\pi}\; {\rm Im}[G({\Bf k},\omega)].
\end{equation}
The energy $\omega_{\Bf k}$ must satisfy $\omega_{\Bf k}=\omega_{\Bf k}^0 
+ \Sigma({\Bf k},\omega_{\Bf k})$, the quasi-particle equation. This 
equation {\sl may or may not} have a solution; however, $\omega_F$ is {\sl 
always} a solution of this equation (see Farid 1999a). If $\omega_{\Bf k}$ 
satisfies this equation and, moreover, $\Sigma({\Bf k},\omega)$ is a 
continuously-differentiable function of $\omega$ in a neighbourhood of 
$\omega=\omega_{\Bf k}$, then one has
\begin{equation}
\label{e11c}
\int_{{\cal C}_{\Bf k}} \frac{{\rm d} z}{2\pi i}\;
G({\Bf k},z) = Z_{\Bf k},
\end{equation}
where
\begin{equation}
\label{e11d}
Z_{\Bf k} {:=} \Big(1 -\left.\frac{\partial 
\Sigma({\Bf k},\omega)}{\partial\omega}\right|_{\omega
=\omega_{\Bf k}}\Big)^{-1}.
\end{equation}
A specific aspect of FLs, in comparison with LLs and marginal FLs, is 
that $\Sigma({\Bf k}_F,\omega)$ pertaining to FLs is a continuously 
differentiable function of $\omega$ in a neighbourhood of $\omega=
\omega_F$ (see text preceding Eq.~(\ref{e1}) above). As will become 
evident, $Z_{{\Bf k}_F} \in (0,1]$, which applies to FLs (specifically, 
but not exclusively) is of special significance (see Eqs.~(\ref{e9}), 
(\ref{e11}) and (\ref{e19a})). Amongst others, by continuity, 
$Z_{{\Bf k}_F}\not=0$ implies a non-vanishing $Z_{\Bf k}$ in a 
neighbourhood of ${\Bf k}={\Bf k}_F$ which together with $\Theta(\mu-
\omega_{\Bf k})$ result in the fact that the first term on the RHS of 
Eq.~(\ref{e11b}) is a non-trivial function which is non-vanishing only 
over a {\sl finite} interval. To appreciate the far-reaching consequence 
of this property in the context of our present discussions, let us consider 
a FL in $d=1$ for which we have
\begin{eqnarray}
\label{e11e}
\int_{-\infty}^{\infty} \frac{{\rm d}k}{2\pi}\;
{\rm e}^{i k x}\, \big\{\Theta(\mu-\omega_k)\, \int_{{\cal C}_k} 
\frac{{\rm d} z}{2\pi i}\,
G(k,z) \big\} &=& \int_{-k_F}^{k_F} \frac{{\rm d} k}{2\pi}\;
{\rm e}^{i k x}\, Z_k\nonumber\\
&=& \frac{Z_{k_F}}{\pi}\,\frac{\sin(k_F x)}{x}\,
- \frac{1}{i x} \int_{-k_F}^{k_F} \frac{{\rm d} k}{2\pi}\;
{\rm e}^{i k x}\; \frac{{\rm d}}{{\rm d} k} Z_k,
\end{eqnarray}
where the second line on the RHS of Eq.~(\ref{e11e}) has been obtained 
through integrating by parts. Under the assumption of integrability of 
${\rm d} Z_k/{\rm d} k$ over $[-k_F,k_F]$, it can be readily demonstrated 
that for $\vert x\vert \to\infty$ the second term on the RHS of 
Eq.~(\ref{e11e}) is asymptotically sub-dominant with respect to the first 
term, whence absence of any contribution due to this second term to the 
leading asymptotic term of $-i G(x,t=0^-)$, $\vert x\vert\to \infty$, as 
presented in Eq.~(\ref{e9}). One can further demonstrate that the 
unbounded support of the second term on the RHS of Eq.~(\ref{e11b}) 
results in the fact that the contribution of this term to the leading 
term in the large-$\vert x\vert$ asymptotic series of $-i G(x,t=0^-)$ is 
vanishing. This simple example brings out the importance of a contribution 
of {\sl finite} support to ${\sf n}({\Bf k})$.

If one applies the same procedure to a case (say, in $d=1$) where the 
first term on the RHS of Eq.~(\ref{e11b}) is identically vanishing (as 
is the case in the one-dimensional Luttinger model), then one readily 
establishes that the leading contribution to $-i G(x,t=0^-)$, for $\vert 
x\vert\to \infty$, decays faster than $1/x$ for $\vert x\vert\to \infty$; 
the expression in Eq.~(\ref{e10}) is an evidence to this statement. On 
the other hand, if $Z_k$ is non-vanishing but ${\rm d} Z_k/{\rm d}k$ is 
non-integrable over $[-k_F, k_F]$ --- a hypothetical situation ---, then 
integration by parts, as applied in Eq.~(\ref{e11e}), cannot be employed 
and an analysis based on some specific aspects of $Z_k$ reveals that 
$-i G(x,t=0^-) \sim o(1)$, for $\vert x\vert\to\infty$, where $o(1)$ 
signifies a vanishingly small function which in the case at hand is 
asymptotically more dominant in comparison with $1/x$.

In dealing with the asymptotic behaviour of $-i G(x,t=0^-)$ for $\vert 
x\vert\to \infty$ in the case of the one-dimensional Luttinger model, one 
has to bear in mind the important fact that here {\sl all} functions of 
momentum are defined over $(-\infty,\infty)$ and that a mathematically 
sound treatment of the model requires introduction of appropriate cut-offs 
(which in considerations concerning asymptotic behaviour of correlation 
functions at large distance must be `soft' cut-offs, as opposed to `sharp' 
ones) in dealing with functions of momentum (Mattis and Lieb 1965). As 
a consequence of this, the boundary contributions obtained through 
integration by parts as applied to the second term on the RHS of 
Eq.~(\ref{e11b}) ({\sl c.f.} the first term on the RHS of Eq.~(\ref{e11e})) 
are vanishing. It is for this very reason that a term decaying like 
$1/x$ is missing in the expression on the RHS of Eq.~(\ref{e10}); this 
missing term, which we denote by $-i G(x,t=0^-)\vert_{\rm missing}$, would 
have the following functional form
\begin{eqnarray}
\label{e11f}
\left. -i G(x,t=0^-)\right|_{\rm missing}
&\equiv& \frac{1}{2\pi i}\,
\frac{\exp(i{\tilde\lambda} x/2) {\sf n}({\tilde\lambda}/2)
-\exp(-i{\tilde\lambda} x/2) {\sf n}(-{\tilde\lambda}/2)}{x}\nonumber\\
&\sim& \frac{-1}{2\pi i}\,
\frac{\exp(-i{\tilde\lambda} x/2)}{x},
\end{eqnarray}
where in the second expression we have employed ${\sf n}(-{\tilde\lambda}/2)
\sim 1$ (due to the assumed absence of a momentum cut-off) and 
${\sf n}({\tilde\lambda}/2) \sim 0$ which apply for sufficiently large 
(positive) ${\tilde\lambda}$. Thus without proper treatment of functions 
of momentum, the distinction between FLs and LLs (here in $d=1$) would be 
obliterated in as far as the decaying behaviour of $-i G(x,t=0^-)$ for 
large $\vert x\vert$ is concerned. The same statement applies for arbitrary 
$d$. Below, by explicitly analysing the large-$\|{\Bf x}\|$ behaviour of 
$-i G({\Bf x},t=0^-)$ for the model under consideration, we demonstrate 
that within the Haldane's (1992) bosonisation scheme for $d > 1$, the 
large-$\|{\Bf x}\|$ behaviour of $-i G({\Bf x},t=0^-)$ {\sl cannot} involve 
an anomalous dimension similar to the case in $d=1$, that is irrespective 
of whether the system under consideration be a FL or otherwise, the 
large-$\|{\Bf x}\|$ behaviour of $-i G({\Bf x},t=0^-)$ is always FL-like.
This suggests an improper treatment of ${\sf n}({\Bf k})$ within this 
scheme when $d > 1$. We establish this fact in Section V.

\subsection{The long-distance behaviour of the Green function within 
the framework of the Haldane bosonisation}

It can be readily shown that
\begin{eqnarray}
\label{e12}
{\rm Re}\left[\frac{2\pi i}{\Omega^2}
{\cal F}^{-1}_{{\Bf x},t=0^-}[\delta G_{\rm b}]\right]
&=& \frac{-2}{(2\pi)^3} \Big\{
\int_0^{2\pi} \frac{{\rm d}\varphi}{2\pi} \int_0^{Q(\varphi)}
{\rm d}\|{\Bf q}\|\;
\cos\big(\|{\Bf q}\|\,\|{\Bf x}\|
\cos(\varphi-\Phi_{{\hat {\Bf x}}})\big)\,\nonumber\\
& &\;\;\;\;\;\;\;\;\;\;\;\;\;\;\;\;\;\;\;\;
\times {\wp}\int_0^1 \frac{{\rm d}\omega}{\pi}\;
\frac{{\rm Im}[W({\Bf q},-\omega)]}{\big(\omega
+\cos(\varphi)\big)^2}\nonumber\\
& &\;\;\;\;\;\;\;\;\;\;
+\int_{\pi/2}^{3\pi/2} \frac{{\rm d}\varphi}{2\pi}\;
\int_0^{Q(\varphi)} {\rm d}\|{\Bf q}\|\;
\cos\big(\|{\Bf q}\|\,\|{\Bf x}\|
\cos(\varphi-\Phi_{{\hat {\Bf x}}})\big)\,\nonumber\\
& &\;\;\;\;\;\;\;\;\;\;\;\;\;\;\;\;\;\;\;\;
\times {\rm Re}[W^{(0,1)}\big(\|{\Bf q}\|,\cos(\varphi)\big)]\Big\},
\end{eqnarray}
where $\Phi_{{\hat {\Bf x}}}$ is the planar angle between ${\Bf x}$ 
and $\mbox{\boldmath{$n$}}_S$ and $Q(\varphi)$ specifies the boundary of 
the rectangular area $[-\lambda_{\perp}/2,\lambda_{\perp}/2] \times 
[-\lambda_{\parallel}/2,\lambda_{\parallel}/2]$ in the momentum space 
over which momentum integrations are carried out (for the origins of
these parameters see HKM (1994) and HKMS (1994)). In general
\footnote{\label{f15}
\small
In both HKM (1994) and HKMS (1994) $\lambda_{\perp}$ is set equal to 
$\lambda$ while $\lambda_{\parallel}$ is equated with $+\infty$. In (Kwon, 
Houghton and Marston 1995 --- see text following Eq.~(27) herein) conditions 
under which $\lambda_{\perp} \ll \lambda_{\parallel}$ or $\lambda_{\perp} 
= \lambda_{\parallel} = \lambda$, etc., apply have been explicitly
indicated (it appears, however, that ``perpendicular directions'' in 
this text should be understood as meaning ``parallel directions''). The 
precise relationship between $\lambda_{\perp}$ and $\lambda_{\parallel}$ 
is of {\sl no} consequence to our observations to be made further on. }
$\lambda_{\perp} \ll \lambda_{\parallel}$. With $\Phi_0 {:=} 
\arctan(\lambda_{\perp}/\lambda_{\parallel})$, we have $Q(\varphi) = 
\lambda_{\parallel}/[2\cos(\varphi)]$, for $0 \leq \varphi < \Phi_0$ 
and $2\pi -\Phi_0 \leq \varphi < 2\pi$; $Q(\varphi) = \lambda_{\perp}/
[2\sin(\varphi)]$, for $\Phi_0 \leq \varphi <\pi -\Phi_0$; $Q(\varphi) = 
-\lambda_{\parallel}/[2\cos(\varphi)]$, for $\pi-\Phi_0 \leq\varphi 
< \pi + \Phi_0$, and $Q(\varphi) = -\lambda_{\perp}/[2\sin(\varphi)]$, 
for $\pi+\Phi_0 \leq\varphi < 2\pi - \Phi_0$. Further, 
$W^{(0,1)}({\Bf q},\omega) {:=} {\rm d} W({\Bf q},\omega)/
{\rm d}\omega$.
After an appropriate regularisation (Farid 1999b), in order to rendering 
it meaningful, it can be rigorously shown that
\begin{equation}
\label{e13}
{\wp}\int_0^1 \frac{{\rm d}\omega}{\pi}\;
\frac{{\rm Im}[W({\Bf q},-\omega)]}{\big(\omega
+\cos(\varphi)\big)^2} \sim
\frac{- N(0) v^2({\Bf q})}{\pi \big(1 + v({\Bf q}) N(0)\big)^2}
\, \ln\vert\cos(\varphi)\vert,\;\;\; \varphi \to \pi/2, 3\pi/2;
\end{equation}
that is to say, for $\varphi\in [0,2\pi]$ the most divergent 
contribution of the integral on the left-hand side (LHS) arises from
neighbourhoods of $\varphi=\pi/2$ and $\varphi=3\pi/2$. Further,
\begin{equation}
\label{e14}
{\rm Re}\big[W^{(0,1)}(\|{\Bf q}\|,\omega)\big]
= \frac{-2 \big(1 + v({\Bf q}) N(0)\big) v^3({\Bf q})
N^2(0)}{d^2 + e^2}\,\omega,\;\;\;\; \vert \omega\vert \leq 1,
\end{equation}
where
\begin{eqnarray}
\label{e15}
d &{:=}& \big(1 + v({\Bf q}) N(0)\big)^2 -
\big[\big(1 + v({\Bf q}) N(0)\big)^2 + 
v^2({\Bf q}) N^2(0)]\big]
\omega^2,\;\;\;\nonumber\\
e &{:=}& 2 \big(1 + v({\Bf q}) N(0) \big) v({\Bf q}) N(0)
\vert\omega\vert (1-\omega^2)^{1/2}.
\end{eqnarray}
For $v({\Bf q})\to\infty$ holds ${\rm Re}\big[W^{(0,1)}(\|{\Bf q}\|,
\omega)\big]=-2\omega/\big[N(0) \big(1+4\omega^2 (1-\omega^2) \big)\big]$.
Evidently, ${\rm Re}\big[W^{(0,1)}(\|{\Bf q}\|,\omega)\big]$
is a regular function of $\omega$ so that the second integral on the RHS
of Eq.~(\ref{e12}) is finite for all $\|{\Bf x}\|$ and therefore from the
view point our present considerations of no significance. Making use of 
the property $\vert\int_a^b {\rm d}x\, f(x)\vert\leq \int_a^b {\rm d}x\, 
\vert f(x)\vert$, applicable to absolutely-integrable functions $f(x)$
over $[a,b]$, it can be readily shown that the RHS of Eq.~(\ref{e12}) 
is bounded for {\sl all} values of $\|{\Bf x}\|$. As a matter of fact, 
explicit analytic calculation (Farid 1999b) shows that for large 
$\|{\Bf x}\|$ both terms on the RHS of Eq.~(\ref{e12}) approach zero. 
For completeness, for the system under consideration to be a LL, it is 
necessary that for $\Phi_{{\hat {\Bf x}}}=0$ the RHS of Eq.~(\ref{e12}) 
behave like $-\alpha \ln\|{\Bf x}\|$ for large $\|{\Bf x}\|$, with 
$\alpha\in (0,1)$.

The above observations are interesting for the reason that they imply 
that {\sl irrespective} of the nature of the interaction function
$v({\Bf q})$, i.e. whether $v({\Bf q})$ be short-, long- or super-long 
range, the metallic state of the system would be a FL (in the narrow 
sense that $Z_{{\Bf k}_F}\not=0$ --- see Farid 1999a). For short- and 
the long-range Coulomb interaction our observation would then agree with 
that by HKM (1994) and HKMS (1994); for super-long ranged interactions, 
however, we would be in conflict with both HKMS (1994) and Bares and 
Wen (1993) who obtain a LL state and a ``$Z_{k_F}=0$ Fermi liquid'', 
respectively. We point out that work by Bares and Wen (1993) is based 
on the random-phase approximation (RPA) so that a discrepancy between 
our finding and that by the latter authors in not inconceivable (see 
however footnote \ref{f13}). 

We have analysed the work by HKMS (1994) and found that the reason for 
their just-mentioned finding lies, amongst others, in their use the 
following approximation (see Eq.~(26) in HKMS (1994))
\begin{equation}
\label{e16}
{\bar\chi}_0(x) = -\frac{N(0)}{2 x^2} + i\eta + {\cal O}(1/x^4),
\end{equation}
which for sufficiently small values of the coupling constant of 
interaction, i.e. $g$ in $v({\Bf q}) = g/\|{\Bf q}\|^{\gamma}$, induces 
spurious poles in, e.g., $W({\Bf q},\omega)$ over a substantial part 
of the ${\Bf q}$-space.
\footnote{\label{f16}
\small
It can be straightforwardly shown that for $\vert\omega\vert \leq 1$,
${\rm Im}\big[W({\Bf q},\omega)\big] = - v^2({\Bf q}) N(0)\,\vert\omega
\vert\, (1-\omega^2)^{1/2}/(a - b \omega^2)$ where $a {:=} \big(1 + 
v({\Bf q}) N(0)\big)^2$ and $b {:=} 1 + 2 v({\Bf q}) N(0)$. Since $a/b > 
1$, it is evident that ${\rm Im}\big[W({\Bf q},\omega)\big]$ is bounded 
for {\sl all} $\vert\omega\vert\in [0,1]$. Employing the approximation 
in Eq.~(\ref{e16}), on the other hand, results in ${\rm Im}\big[W({\Bf q},
\omega)\big] = -(\pi/2)\,{\rm sgn}(\omega)\,\sqrt{v({\Bf q}) N(0)/2}\, 
\delta\big(\omega -{\rm sgn}(\omega)\,\sqrt{v({\Bf q}) N(0)/2}\big)$. 
This implies that $W({\Bf q},\omega)$ has simple poles at $\omega
=\omega_{\pm} {:=} \pm \sqrt{v({\Bf q}) N(0)/2}$ which for sufficiently 
large $\|{\Bf q}\|$, or sufficiently small $g$ {\sl and} over a substantial 
part of the relevant ${\Bf q}$-space, are located in the interval $(-1,1)$. 
With the plasma frequency $\omega_{\rm p}$ obeying $\omega_{\rm p} = \big(g 
\|{\Bf k}_F\|/[2\pi]\big)^{1/2}$, it is interesting to mention that HKMS 
(1994) assert that ``If $\omega_{\rm p} \gg \lambda$ the system does not 
see the plasmons and Fermi liquid behaviour is retained. In the opposite 
limit of $\omega_{\rm p} \ll \lambda$, however, the plasmons destroy the 
Fermi liquid.'' The property $\omega_{\rm p}\propto \vert\omega_{\pm}\vert$ 
leaves no doubt that in the case at hand destruction of the FL state 
far from being caused by plasmons is due to failure of the approximate 
expression in Eq.~(\ref{e16}) to be qualitatively correct.}
Thus we seem to have no alternative but to accept the verdict stated at 
the beginning of the previous paragraph.

\subsection{Investigating some crucial aspects}

In view of the above conflicting findings, we now turn to investigating
in some detail a number of crucial aspects in Haldane's (1992) 
bosonisation scheme. As the point of departure, we consider the case of 
non-interacting fermions and determine the behaviour of $G_0({\Bf x},
t=0^-)$ as evaluated from the expression in Eq.~(\ref{e3b}). This
will lead us to a key observation concerning a shortcoming of the 
operator algebra (see Eq.~(\ref{e28}) below) that underlies 
Eq.~(\ref{e3}). We have
\footnote{\label{f17}
\small
It is readily verified that $\int_{-\infty}^{\infty} {\rm d}x\,
\exp(- i k x) \int_{-\infty}^{\infty} {\rm d}t\, \exp(i \omega t)/
\big[2\pi \big(x - t + i\,\eta\,{\rm sgn}(t)\big)\big] = 1/\big(\omega 
- k + i\,\eta\,{\rm sgn}(\omega)\big) \equiv G_0(k,\omega)$, where 
$\eta\downarrow 0$. Thus the additional pre-factor in Eq.~(\ref{e17}), 
namely $\Lambda/(2\pi)$, is specific to the `patch' Green function which, 
as we shall see, is crucial for an averaging procedure. } 
\begin{equation}
\label{e17}
G_{0;{\rm f}}({\Bf x},t=0^-;S) = \frac{\Lambda}{(2\pi)^2}\,
\frac{\exp\big(i {\Bf k}_F(S)\cdot {\Bf x}\big)}{{\Bf x}\cdot
\mbox{\boldmath{$n$}}_S - i\,\eta},\;\;\;\eta\downarrow 0.
\end{equation}

We obtain the non-interacting Green function $G_0({\Bf x},t=0^-)$
through averaging (below indicated by `Av') the RHS of Eq.~(\ref{e17}) 
over all `patches' of the FS (for some subtle aspects related to this 
see further on). In doing so, we make use of $\Lambda \ll 1$ and thus 
employ the following result which is applicable to the limit $\Lambda
\downarrow 0$ (see text following Eq.~(\ref{e4}) above),
\begin{equation}
\label{e17a}
\sum_{S}\, (\dots) = \frac{\|{\Bf k}_F\|}{\Lambda}
\int_0^{2\pi} {\rm d}\varphi\; (\dots).
\end{equation}
With ${\Bf k}_F(S) = \|{\Bf k}_F\|\, \mbox{\boldmath{$n$}}_S$, where we 
assume $\|{\Bf k}_F\|$ to be independent of patch index, and ${\Bf x}
\cdot \mbox{\boldmath{$n$}}_S = \|{\Bf x}\|\,\cos(\varphi)$, we have
\begin{eqnarray}
\label{e18}
-i G_0^{\rm Av}({\Bf x},t=0^-)
&=& \frac{\|{\Bf k}_F\|}{\|{\Bf x}\|} \int_0^{2\pi} 
\frac{{\rm d}\varphi}{2\pi}\;
{\rm e}^{i \|{\Bf k}_F\|\,\|{\Bf x}\| \cos(\varphi)}
\, \delta\big(\cos(\varphi)\big)\nonumber\\
& &\;\;\;\;\;\;
-\frac{i\,\|{\Bf k}_F\|}{2\pi \|{\Bf x}\|}\,
{\wp}\int_0^{2\pi} \frac{{\rm d}\varphi}{2\pi}\;
\frac{\exp\big(i \|{\Bf k}_F\|\,
\|{\Bf x}\|\cos(\varphi)\big)}{\cos(\varphi)},
\end{eqnarray}
which after some algebra results in
\begin{equation}
\label{e19}
-i G_0^{\rm Av}({\Bf x},t=0^-) \sim \frac{\|{\Bf k}_F\|}{2\pi\,
\|{\Bf x}\|}\, 
+ \frac{\|{\Bf k}_F\|^{1/2}}{2^{1/2} \pi^{3/2}}
\, \frac{\sin(\|{\Bf k}_F\|\,\|{\Bf x}\|
-\pi/4)}{\|{\Bf x}\|^{3/2}},\;\;\; \|{\Bf x}\|\to\infty,
\end{equation}
of which the second term on the RHS is obtained through application of the 
method of stationary phase (see, e.g., Ch.~4 in Murray 1974), taking into 
account that ${\rm d}\cos(\varphi)/{\rm d}\varphi\equiv -\sin(\varphi) 
= 0$ for $\varphi=0, \pi, 2\pi$ with $\varphi =0, 2\pi$ coinciding with 
the end-points of the interval of integration (see pp.~76 and 77 in Murray 
1974). Since a system of non-interacting fermions is a FL, this expression 
must be compared with that in Eq.~(\ref{e11}). 

The first term on the RHS of Eq.~(\ref{e19}) is not unexpectedly 
spurious: it is a contribution from points $\varphi=\pi/2, 3\pi/2$ at 
which for large $\|{\Bf x}\|$ the condition for validity of Eq.~(\ref{e3})
and therefore of Eq.~(\ref{e3b}), namely $\vert {\Bf x}\cdot 
\mbox{\boldmath{$\tau$}}_S\vert\ll 1/\Lambda$, is violated. 

The second term on the RHS of Eq.~(\ref{e19}), contrary to the first 
term, cannot be suffering from violation of condition $\vert {\Bf x}\cdot 
\mbox{\boldmath{$\tau$}}_S\vert \ll 1/\Lambda$ since it entirely consists 
of contributions from the turning points of $\cos(\varphi)$, namely 
$\varphi=0, \pi, 2\pi$ (see above), and at these points we have $\vert 
{\Bf x}\cdot \mbox{\boldmath{$\tau$}}_S\vert =0$ for {\sl all} ${\Bf x}$. 
Having established this fact, it follows that the perfect agreement 
between this term and the RHS of Eq.~(\ref{e11}) with $Z_{{\Bf k}_F}=1$ 
is not accidental.

For our following considerations we introduce ${\cal F}^{-1}_{(\|{\Bf x}\|,
\varphi),t}[\delta G_{\rm b}]$ as representing ${\cal F}^{-1}_{{\Bf x},
t}[\delta G_{\rm b}]$ in the cylindrical coordinates. Since 
${\cal F}^{-1}_{(\|{\Bf x}\|,\varphi),0^-}[\delta G_{\rm b}] \equiv 
{\cal F}^{-1}_{(\|{\Bf x}\|,\varphi + m\pi),0^-}[\delta G_{\rm b}]$ 
with $m$ any integer (Farid 1999b), it can be readily shown that
\begin{eqnarray}
\label{e19a}
- i G^{\rm Av}({\Bf x},t=0^-) &\sim&
\frac{\Xi\, \|{\Bf k}_F\|}{2\pi \|{\Bf x}\|}\,
\exp\big(2\pi i\Omega^{-2} {\cal F}^{-1}_{(\|{\Bf x}\|,\pi/2),t=0^-}
[\delta G_{\rm b}]\big)\nonumber\\
& &\;\; + \frac{\Xi\, \|{\Bf k}_F\|^{1/2}}{2^{1/2} \pi^{3/2}}\,
\exp\big(2\pi i\Omega^{-2} {\cal F}^{-1}_{(\|{\Bf x}\|,0),t=0^-}
[\delta G_{\rm b}]\big)
\nonumber\\
& &\;\;\;\;\;\times \frac{\sin\big(\|{\Bf k}_F\|\, 
\|{\Bf x}\| -\pi/4\big)}{\|{\Bf 
x}\|^{3/2}},\;\;\; \|{\Bf x}\| \to \infty.
\end{eqnarray}
All our remarks concerning the nature of the two terms on the RHS of 
Eq.~(\ref{e19}) apply to those on the RHS of Eq.~(\ref{e19a}). 

Since, as we have mentioned above (see text following Eq.~(\ref{e15}) 
above), $\exp\big(2\pi i\Omega^{-2} {\cal F}^{-1}_{(\|{\Bf x}\|,0),t=0^-}
[\delta G_{\rm b}]\big)\to 1$ for $\|{\Bf x}\|\to\infty$, independent 
of the nature of the particle-particle interaction, it follows that we 
can directly compare the second term on the RHS of Eq.~(\ref{e19a}) with 
the term on the RHS of Eq.~(\ref{e11}), and thus identify
\footnote{\label{f18}
\small
For completeness, if $\exp\big(2\pi i\Omega^{-2}{\cal F}^{-1}_{(\|
{\Bf x}\|,0), t=0^-}[\delta G_{\rm b}]\big)\to C$ for $\|{\Bf x}\|
\to\infty$, then we would have $C \Xi = Z_{{\Bf k}_F}$. If, on the 
other hand, $2\pi i\Omega^{-2} {\cal F}^{-1}_{(\|{\Bf x}\|,0),t=0^-}
[\delta G_{\rm b}]\to \infty$ for $\|{\Bf x}\|\to\infty$, Eq.~(\ref{e19a})
{\sl cannot} be compared with Eq.~(\ref{e11}) and therefore $\Xi$
{\sl cannot} be identified with $Z_{{\Bf k}_F}$. Recall that (see text 
following Eq.~(\ref{e3a}) above) $\Xi\not=0$, for otherwise $G_{\rm f} 
\equiv 0$.}
$\Xi$ with $Z_{{\Bf k}_F}$. This simple observation leads us to draw a 
far-reaching conclusion: in light of our discussions following 
Eq.~(\ref{e3d}) and with reference to Eq.~(\ref{e7}), $\Xi = 
Z_{{\Bf k}_F}$ {\sl and} $\Sigma(0)=0$. Moreover, our finding that the 
real part of the second term on the RHS of Eq.~(\ref{e3c}) does {\sl not} 
scale like $1/\omega$, as $\vert\omega\vert\to 0$ (see paragraph following 
Eq.~(\ref{e7}) above), is {\sl not} an artifact of the first-order 
expansion employed by us (although our result that this contribution 
scales like $\ln^2\vert\omega\vert$ may best be). 

% V
\section{The algebraic structure of the Haldane bosonisation scheme
revisited} 

In order to gain insight into the origin of the above results, we proceed 
with an investigation of the algebraic structure underlying Hadalne's (1992) 
bosonisation scheme which is embodied by a Kac-Moody algebra. In deriving 
this algebra for the charge-current operators ${\hat J}(S;{\Bf q})$ and 
${\hat J}(T;{\Bf q})$ pertaining to FS `patches' $S$ and $T$, 
\begin{equation}
\label{e21}
{\hat J}(S;{\Bf q}) {:=} \sum_{\Bf k\,\sigma} \Theta(S;{\Bf k}+{\Bf q})
\Theta(S;{\Bf k}) \Big\{{\hat a}_{{\Bf k}+{\Bf q}\,\sigma}^{\dag}
{\hat a}_{{\Bf k}\,\sigma} - \delta_{{\Bf q},{\Bf 0}}\,
{\sf n}({\Bf k}) \Big\},
\end{equation}
one starts from the following {\sl exact} result (below $[\,,\,]_{-}$ 
denotes commutation)
\begin{eqnarray}
\label{e22}
[{\hat J}(S;{\Bf q}),{\hat J}(T;{\Bf p})]_- &=& \sum_{{\Bf k}\,\sigma}
\Big\{ \Theta(S;{\Bf k}+{\Bf p}+ {\Bf q})\,\Theta(S;{\Bf k}+{\Bf p})\,
\Theta(T;{\Bf k}+{\Bf p})\,\Theta(T;{\Bf k})\nonumber\\
& &\mbox{} -\Theta(S;{\Bf k}+{\Bf q})\,
\Theta(S;{\Bf k})\,\Theta(T;{\Bf k}+{\Bf p}+{\Bf q})\,
\Theta(T;{\Bf k}+{\Bf q})\Big\}\,
{\hat a}_{{\Bf k}+{\Bf p}+{\Bf q}\,\sigma}^{\dag}
{\hat a}_{{\Bf k}\,\sigma}
\end{eqnarray}
which is obtained through repeated application of the canonical
anti-commutation relations that hold amongst the fermion creation and 
annihilation operators ${\hat a}^{\dag}_{{\Bf k}\,\sigma}$ and 
${\hat a}_{{\Bf k}\,\sigma}$, respectively; here $\sigma$ denotes the 
spin index. Above $\Theta(S;{\Bf k})$ stands for the characteristic 
function of the FS `patch' $S$, i.e. it is equal to $1$ when ${\Bf k}$ 
lies inside the `squat box' associated with `patch' $S$ and zero 
otherwise. {\sl Two} fundamental assumptions (the second of which to our 
best knowledge has never earlier been acknowledged as such) are necessary 
in order to reduce the expression in Eq.~(\ref{e22}) into a Kac-Moody 
algebra. The first of these amounts to a decoupling of scattering processes 
amongst different `squat boxes' and is expressed through the following 
two equations:
\begin{eqnarray}
\label{e23}
\Theta(S;{\Bf k}+{\Bf p}+{\Bf q})\,\Theta(T;{\Bf k}+{\Bf p})
&=& \delta_{S,T}\, \Theta(S;{\Bf k}+{\Bf p}+{\Bf q})\,\Theta(S;{\Bf k}
+{\Bf p}),\\
\label{e24}
\Theta(S;{\Bf k})\,\Theta(T;{\Bf k}+{\Bf q})
&=& \delta_{S,T}\,\Theta(S;{\Bf k})\, \Theta(S;{\Bf k}+{\Bf q}).
\end{eqnarray}
In order to expose the second {\sl approximation}, we first need to proceed 
through two intermediate steps. In the first, $\delta_{{\Bf q}+{\Bf p},
{\Bf 0}}\,{\sf n}({\Bf k})$ is added and subsequently subtracted from 
$\sum_{\sigma} {\hat a}^{\dag}_{{\Bf k}+{\Bf p}+{\Bf q}\,\sigma}
{\hat a}_{{\Bf k}\,\sigma}$ in Eq.~(\ref{e22}), taking into account
the assumptions in Eqs.~(\ref{e23}) and (\ref{e24}). In the second, on 
the basis of the inequalities $\lambda\ll \Lambda \ll \|{\Bf k}_F\|$ an 
argument is made (see Houghton and Marston 1993 and Haldane 1992) to 
write the thus-obtained simplified expression as follows
\begin{eqnarray}
\label{e25}
[{\hat J}(S;{\Bf q}),{\hat J}(T;{\Bf p})]_- &=&
\delta_{S,T}\,\Big\{
\sum_{\Bf k}\,\Theta(S;{\Bf k}+{\Bf p}
+{\Bf q})\,\Theta(S;{\Bf k})\,\big[\Theta(S;{\Bf k}+{\Bf p})
-\Theta(S;{\Bf k}+{\Bf q})\big]\nonumber\\
& &\;\;\;\;\;\times
\delta_{{\Bf q}+{\Bf p},{\Bf 0}}\, {\sf n}({\Bf k})
+ \mbox{\rm some error term} \Big\}.
\end{eqnarray}
Owing to $\delta_{{\Bf q}+{\Bf p},{\Bf 0}}$, ${\Bf p}$ in the summand can 
be replaced by $-{\Bf q}$, upon which, following a shift in the summation 
variable, the RHS of Eq.~(\ref{e25}) in the thermodynamic limit transforms
into an expression involving
\begin{equation}
\label{e26}
g_{S}({\Bf q}) {:=} \int_{{\bar S}} {\rm d}^d{\Bf k}\; \big[
{\sf n}({\Bf k}-{\Bf q}/2) - {\sf n}({\Bf k}+{\Bf q}/2)\big],
\end{equation}
in which ${\bar S}$ denotes the `squat box' defined by the region of the
${\Bf k}$-space for which holds $\Theta(S;{\Bf k}+{\Bf q}/2)
\Theta(S;{\Bf k}-{\Bf q}/2) = 1$. Now as for the second approximation
to which we have referred above, Haldane (1992)
\footnote{\label{f19}
\small
See p.~20 in (Haldane 1992). The function $g_{S}({\Bf q})$
as defined in Eq.~(\ref{e26}) is identical to $g_{\alpha}
(\mbox{\boldmath{$q$}})$ as defined in Eq.~(35) in (Haldane 1992),
which, corrected for some misprints, reads: 
$g_{\alpha}(\mbox{\boldmath{$q$}}) = \sum_{\mbox{\boldmath{$k$}}}
\theta_{\alpha}(\mbox{\boldmath{$k$}} + \mbox{\boldmath{$q$}})\, 
\theta_{\alpha}(\mbox{\boldmath{$k$}})
\,\langle (n_{\mbox{\boldmath{$k$}} + \mbox{\boldmath{$q$}}} - 
n_{\mbox{\boldmath{$k$}}}) \rangle_0$.
It is perhaps useful to mention that $a$ and $V$ in Eq.~(36) of 
(Haldane 1992) are defined as $a {:=} \Lambda^{d-1}$ and $V {:=} 
2 (L/[2\pi])^d$. Further, $\Lambda$ and $\lambda$ in the work
by Haldane (1992) are to be identified with respectively $\lambda$ and 
$\Lambda$ in the work by Houghton and Marston (1993) whose notation
we have adopted in our present work. } 
asserts that the value of $g_{S}({\Bf q})$, for $\|{\Bf q}\| \ll\lambda$, 
which is ``the number of allowed $k$-space points inside the volume of 
reciprocal space swept out by displacing the patch of Fermi surface by 
$\mbox{\boldmath{$q$}}$'', ``is independent of the detailed structure of 
$\langle n_{\mbox{\boldmath{$k$}}} \rangle_0$ [$\equiv {\sf n}({\Bf k})$] 
near the Fermi surface and only involves the change in the asymptotic 
values of the occupation factor from deep inside to far outside the Fermi 
surface.'' This statement, as we shall explicitly demonstrate below, is 
applicable {\sl only} in $d=1$. Moreover, we unequivocally demonstrate 
that in $d > 1$, specifically in $d=2$, the employed Kac-Moody algebra 
for current operators does {\sl not} allow $Z_{{\Bf k}_F}$ to be vanishing
and the large-$\|{\Bf x}\|$ behaviour of $-i G({\Bf x},t=0^-)$ to be 
different from that of FLs, in full conformity with our finding based on our 
explicit calculation of $-i G({\Bf x},t=0^-)$ for $\|{\Bf x}\|\to \infty$.  

It is on the basis of the view point of Haldane's, leading to
\begin{equation}
\label{e27}
g_{S}({\Bf q}) = \Lambda^{d-1}\, {\Bf q}\cdot
\mbox{\boldmath{$n$}}_S,
\end{equation}
that Eq.~(\ref{e25}) upon suppression of `some error term' furnishes 
the charge-current operators $\{{\hat J}(S;{\Bf q})\}$ with the Kac-Moody 
current algebra (Brink and Hanneaux 1988, Goddard and Olive 1986)
\begin{equation}
\label{e28}
\Big[{\hat J}(S;{\Bf q}), {\hat J}(T;{\Bf p})\Big]_{-} =
2 \delta_{S,T} \delta_{{\Bf q}+{\Bf p},{\Bf 0}}\,\Omega
\,{\Bf q}\cdot \mbox{\boldmath{$n$}}_S,
\end{equation}
where $\Omega {:=} \Lambda^{d-1} (L/[2\pi])^d$, with $L$ the length of 
the side of the $d$-dimensional macroscopic hyper-cube within which the 
system under consideration is confined ($\Lambda$ is defined in the 
paragraph containing Eq.~(\ref{e3}) above). For later reference, we 
mention that the RHS of Eq.~(\ref{e28}) is known as {\sl quantum anomaly}. 

Now we establish that in a rigorous treatment of the problem at hand, 
`some error term' in Eq.~(\ref{e25}) and/or the `inter-box' scattering 
events, as expressed in Eqs.~(\ref{e23}) and (\ref{e24}), must play 
prominent roles.
\footnote{\label{f20}
\small
This `some error term' is in the work by Haldane (1992) --- Eq.~(34) herein 
--- denoted by $X_{\alpha}(\mbox{\boldmath{$q$}},\mbox{\boldmath{$q$}}')$.
Haldane remarks (see p.~20 in Haldane 1992): ``Because (in contrast to 
the original Tomonaga calculation in 1D) some of this 
$\prec\!\!\prec$correction$\succ\!\!\succ$ involves states at the Fermi 
surface, this is perhaps not as innocuous an approximation in higher 
dimensions, but appears to be valid in the long-wavelength limit.'' 
In light of our considerations, it should be evident that ``but appears 
to be valid in the long-wavelength limit'' is not tenable. Note that full 
inclusion of  $X_{\alpha}(\mbox{\boldmath{$q$}}, \mbox{\boldmath{$q$}}')$ 
destroys the Kac-Moody algebra. }
In this connection we mention that the estimation as put forward by 
Houghton and Marston (1993) --- see text following Eq.~(3.5a) herein;
the authors explicitly consider the case of $d=3$ ---, namely that the 
magnitude of `some error term' were by on the order of $1/\sqrt{{\cal N}}$ 
smaller than that of the quantum anomaly, 
\footnote{\label{f21}
\small
Houghton and Marston (1993) estimate (specifically for $d=3$) the size 
of `some error term' to be on the order of $(\Lambda^{d-1}\lambda/{\cal N}) 
(L/[2\pi])^d$, to be compared with that of the quantum anomaly which is 
on the order of $(\Lambda^d/{\cal N}) (L/[2\pi])^d$. The choice $\lambda 
= \Lambda/\sqrt{{\cal N}}$ leads to the above-mentioned statement. }
where ${\cal N} {:=} \|{\Bf k}_F\|/\lambda$, aside from being based on the 
approximation in which $\big({\hat a}^{\dag}_{{\Bf k}+{\Bf p}+{\Bf q}\,
\sigma} {\hat a}_{{\Bf k}\,\sigma}-\delta_{{\Bf q} +{\Bf p},{\Bf 0}}\,{\sf n}
({\Bf k})\big)$ is replaced by the $c$-number $(1 - \delta_{{\Bf q}+{\Bf 
p},{\Bf 0}})$, relies on the assumption of validity of the RHS of
Eq.~(\ref{e28}). Below we explicitly deal with cases $d=1$ and $d=2$
and demonstrate that the restrictions $\lambda \ll \Lambda \ll
\|{\Bf k}_F\|$ in $d > 1$ render the Haldane (1992) bosonisation scheme 
in $d > 1$ of limited validity.

% V.a
\subsection{The case of $d=1$}

Here by choosing a positive direction, we consider one-dimensional
{\sl vectors} ${\Bf q}$ and ${\Bf k}$ as {\sl scalars} and define
$k_F {:=} \|{\Bf k}_F\|$. We specifically consider the one-dimensional 
Luttinger model (Luttinger 1963, Mattis and Lieb 1965; see, e.g.,
Voit 1994) for spin-less fermions and in doing so replace ${S}$ by 
$r=\pm$ to distinguish $g_{S}({\Bf q}) \equiv g_r({\Bf q})$ 
associated with the branch of left ($r=-$) and right ($r=+$) movers. 
We have
\begin{equation}
\label{e29}
g_r({\Bf q}) {:=} \int_{-\lambda/2 + r k_F + {\Bf q}/2}^{\lambda/2 +
r k_F - {\Bf q}/2} {\rm d}{\Bf k}\; \big[{\sf n}_r({\Bf k}-{\Bf q}/2)
- {\sf n}_r({\Bf k}+{\Bf q}/2)\big],
\end{equation}
where $\lambda$ is a cut-off parameter (see text preceding Eq.~(\ref{e3})
above). One readily verifies that the choice of the boundaries of the 
integral on the RHS of Eq.~(\ref{e29}) indeed restricts the domain of 
integration to ${\bar S}$.

Through decomposing the integral on the RHS of Eq.~(\ref{e29}) into two 
integrals involving separately ${\sf n}_r({\Bf k}-{\Bf q}/2)$ and 
${\sf n}_r({\Bf k}+{\Bf q}/2)$ and subsequently transforming the variable 
of integration ${\Bf k}$ (${\Bf k}\rightharpoonup {\Bf k}+{\Bf q}/2$ in 
the first and ${\Bf k}\rightharpoonup {\Bf k}-{\Bf q}/2$ in the second), 
cancellation of some contributions to the resulting integrals gives rise 
to $\int_{-\lambda/2+ r k_F}^{-\lambda/2 + r k_F + {\Bf q}} {\rm d}{\Bf k}\,
{\sf n}_r({\Bf k}) - \int_{\lambda/2+ r k_F-{\Bf q}}^{\lambda/2+ r k_F}
{\rm d}{\Bf k}\, {\sf n}_r({\Bf k})$. For sufficiently large positive
$\lambda$ (see text following Eq.~(\ref{e3}) above), ${\sf n}_r({\Bf k})$
in the first integral can be replaced by unity (zero) and in the second
integral by zero (unity) for $r=+$ ($r=-$), thus in combination yielding
\begin{equation}
\label{e30}
g_r({\Bf q}) = r\,{\Bf q}.
\end{equation}
This result is in full conformity with that in Eq.~(\ref{e27}). It 
is important that Eq.~(\ref{e30}), which applies for `sufficiently' 
large $\lambda$, is independent of $\lambda$ itself. Since a correct 
treatment of the problem at hand requires that ${\sf n}_r({\Bf k})$ be 
cut off (in a `soft' manner) for $\vert {\Bf k} \vert$ larger than a 
certain cut-off value, say ${\tilde\lambda}_r > 0$ (see Eq.~(\ref{e11e}) 
above and text following it), it is however necessary that $\vert 
-\lambda/2+r k_F\vert \ll {\tilde\lambda}_r$ ($\vert \lambda/2 + r k_F\vert 
\ll {\tilde\lambda}_r$) for $r=+$ ($r=-$). Since ${\tilde\lambda}_r$ can 
be chosen arbitrarily large, we observe that in $d=1$ nothing stands
in the way of choosing $\lambda$ `sufficiently' large and thus fulfilling
Eq.~(\ref{e30}).

% V.b
\subsection{The case of $d=2$}

In the cylindrical polar coordinates $(\|{\Bf k}\|,\theta)$, with 
${\Bf q}$ the polar axis (i.e. with ${\Bf k}\cdot {\Bf q} = 
\|{\Bf k}\|\,\|{\Bf q}\| \cos\theta$; for convenience we choose the 
origin of ${\Bf k}$ to coincide with the centre of the Fermi sea) 
we have
\begin{equation}
\label{e31}
g_{S}({\Bf q}) = \int_{\theta_{S;1}}^{\theta_{S;2}}
{\rm d}\theta\; \big\{ j_{\Bf q}^{-}(\theta)
- j_{\Bf q}^{+}(\theta)\big\},
\end{equation}
where
\begin{equation}
\label{e32}
j_{\Bf q}^{\pm}(\theta) {:=} \int_{k_1(\theta)}^{k_2(\theta)}
{\rm d}\|{\Bf k}\|\; \|{\Bf k}\|\, {\sf n}\big({\Bf k}(\theta)\pm 
{\Bf q}/2\big).
\end{equation}
In Eq.~(\ref{e31}), $\theta_{S;1}$ and $\theta_{S;2}$, $\theta_{S;1} 
< \theta_{S;2}$, specify the interval $(\theta_{S;1},\theta_{S;2})$ 
associated with `patch' $S$ and in Eq.~(\ref{e32}) $k_1(\theta)$ and 
$k_2(\theta)$, with $\theta\in (\theta_{S;1},\theta_{S;2})$, specify 
the radial extent of `squat box' ${\bar S}$ in the direction $\theta$; 
we have $k_2(\theta) - k_1(\theta) = \lambda$ (see text preceding 
Eq.~(\ref{e3}) above). Making use of $\|{\Bf k}\pm {\Bf q}/2\| \approx 
\|{\Bf k}\| \pm \frac{1}{2} \|{\Bf q}\| \cos\theta$, which is appropriate 
for small $\|{\Bf q}\|$, and some change of variables we obtain
\begin{equation}
\label{e33}
j_{\Bf q}^{\pm}(\theta)
\approx
\int_{k_1(\theta)\pm\frac{1}{2} \|{\Bf q}\| \cos\theta}^{k_2(\theta)
\pm \frac{1}{2} \|{\Bf q}\| \cos\theta} {\rm d} \|{\Bf k}\|\;
\|{\Bf k}\|\, {\sf n}\big({\Bf k}(\theta)\big)
\mp \frac{1}{2} \|{\Bf q}\| \cos\theta
\int_{k_1(\theta)\pm\frac{1}{2} \|{\Bf q}\| \cos\theta}^{k_2(\theta)
\pm \frac{1}{2} \|{\Bf q}\| \cos\theta} {\rm d} \|{\Bf k}\|\; 
{\sf n}\big({\Bf k}(\theta)\big).
\end{equation}
For generality we here take ${\Bf k}_F$ to depend upon $\theta$. With 
$k_1(\theta) < \|{\Bf k}_F(\theta)\| < k_2(\theta)$ we assume
${\sf n}\big({\Bf k}(\theta)\big)$ to be regular in the neighbourhoods 
of $k_1(\theta)$ and $k_2(\theta)$. For sufficiently wide neighbourhoods 
which accommodate $k_j(\theta) \pm\frac{1}{2} \|{\Bf q}\| \cos\theta$, 
$j=1,2$, for given values of $\|{\Bf q}\|$ and $\theta$, we have
\begin{eqnarray}
\label{e34}
& &\int_{k_1(\theta)\pm\frac{1}{2} \|{\Bf q}\| \cos\theta}^{k_2(\theta)
\pm\frac{1}{2} \|{\Bf q}\| \cos\theta} {\rm d} \|{\Bf k}\|\; 
\|{\Bf k}\|\, {\sf n}\big({\Bf k}(\theta)\big) =
\int_{k_1(\theta)}^{k_2(\theta)} {\rm d} \|{\Bf k}\|\; \|{\Bf k}\|\,
{\sf n}\big({\Bf k}(\theta)\big)\nonumber\\
& &\;\;\;\;\;\;\;\;\;\;\;\;\;\;\;\;\;\;\;\;\;\;\;
\pm\frac{1}{2}
\|{\Bf q}\| \cos\theta\, \Big\{
k_2(\theta) {\sf n}\big({\Bf k}_2(\theta)\big) -
k_1(\theta) {\sf n}\big({\Bf k}_1(\theta)\big) \Big\}
+ {\cal O}(\|{\Bf q}\|^2),
\end{eqnarray}
from which we obtain
\begin{equation}
\label{e35}
j_{\Bf q}^-(\theta) - j_{\Bf q}^+(\theta)
= \|{\Bf q}\| \cos\theta\, \Big\{
k_1(\theta) {\sf n}\big({\Bf k}_1(\theta)\big) -
k_2(\theta) {\sf n}\big({\Bf k}_2(\theta)\big) +
\int_{k_1(\theta)}^{k_2(\theta)} {\rm d} \|{\Bf k}\|\;
{\sf n}\big({\Bf k}(\theta)\big) \Big\} + {\cal O}(\|{\Bf q}\|^2).
\end{equation}
The last term on the RHS enclosed by braces makes explicit that in 
$d=2$ (in fact in any $d > 1$) it is not only ${\sf n}({\Bf k})$ far 
inside and far outside the FS that is relevant but that the behaviour 
of ${\sf n}({\Bf k})$ over the entire `squat box' is significant.

Owing to the continuity of the RHS of Eq.~(\ref{e35}), as a function
of $\theta$, the first mean-value theorem (Whittaker and Watson 1927,
p.~65) applies and we can write
\begin{equation}
\label{e36}
g_{S}({\Bf q})
= \big(\theta_{S;2} -\theta_{S;1}\big)\, \Big\{
j_{\Bf q}^-({\tilde\theta}_{S}) -
j_{\Bf q}^+({\tilde\theta}_{S}) \Big\},\;\;\;
\mbox{\rm for {\sl some}}\;\; {\tilde\theta}_{S}\in
[\theta_{S;1},\theta_{S;2}].
\end{equation}
For the sake of argument let us now make an appeal to `common wisdom' and 
make use of the trapezoidal rule (Abramowitz and Stegun 1972, p.~885)
\begin{equation}
\label{e37}
\int_a^{a+\Delta} {\rm d}x\; f(x) = \frac{\Delta}{2}
\big[ f(a+\Delta) + f(a)\big] + {\cal O}(\vert\Delta\vert^3);
\end{equation}
neglecting the rest term we have
\begin{equation}
\label{e38}
\int_{k_1({\tilde\theta}_{S})}^{k_2({\tilde\theta}_{S})}
{\rm d} \|{\Bf k}\|\; {\sf n}\big({\Bf k}({\tilde\theta}_{S})\big)
\approx \frac{1}{2} \big(k_2({\tilde\theta}_{S}) -
k_1({\tilde\theta}_{S}) \big)\,\big[
{\sf n}\big({\Bf k}_2({\tilde\theta}_{S})\big) +
{\sf n}\big({\Bf k}_1({\tilde\theta}_{S})\big) \big].
\end{equation}
Making use of the properties
\footnote{\label{f22}
\small
For interacting particles, ${\sf n}({\Bf k})$ can significantly differ 
from unity (zero) even for ${\Bf k}$  far inside (outside) the Fermi 
sea (for some recent quantum-Monte-Carlo based results concerning 
${\sf n}({\Bf k})$ see Moroni, Senatore and Fantoni 1997). Therefore, it 
would be advisable to leave ${\sf n}({\Bf k}_1)$ and ${\sf n}({\Bf k}_2)$ 
in the formalism and determine them in a self-consistent manner. In 
this connection note that in applying the Haldane (1992) bosonisation 
scheme, the inequalities $\lambda \ll \Lambda \ll \|{\Bf k}_F\|$ are to 
be satisfied and a small $\lambda$ does not allow ${\Bf k}$ to be either 
far inside or far outside the Fermi sea. }
${\sf n}\big({\Bf k}_1({\tilde\theta}_{S}) \big) = 1$ and ${\sf n} 
\big({\Bf k}_2({\tilde\theta}_{S})\big) = 0$ (see texts following 
Eqs.~(\ref{e26}) and (\ref{e29}) above), from Eq.~(\ref{e38}) we obtain
\begin{equation}
\label{e39}
g_{S}({\Bf q}) \approx
\frac{1}{2} \|{\Bf q}\| \cos({\tilde\theta}_{S}) \big(
\theta_{S;2} - \theta_{S;1}\big)\,
\big[ k_1({\tilde\theta}_{S}) + k_2({\tilde\theta}_{S})
\big] = \Lambda\, {\Bf q}\cdot \mbox{\boldmath{$n$}}_{S},
\end{equation}
where we have made use of the fact that $\big[k_1({\tilde\theta}_{S}) 
+ k_2({\tilde\theta}_{S})\big]/2 {=:} {\bar k}({\tilde\theta}_{S})$ 
describes the distance from the centre of the Fermi sea to `patch' $S$ 
of the FS at angel ${\tilde\theta}_S$ (for ${\tilde\theta}_S = (\theta_{S;1}
+\theta_{S;2})/2$, ${\bar k}({\tilde\theta}_S) =\|{\Bf k}_F\|$) and that 
for small $(\theta_{S;2} -\theta_{S;1})$, ${\bar k} ({\tilde\theta}_{S}) 
\times (\theta_{S;2} -\theta_{S;1}) = \Lambda$, the length of the `patch' 
(see text preceding Eq.~(\ref{e3}) above). Further, we have employed
$\|{\Bf q}\| \cos({\tilde\theta}_S) = {\Bf q}\cdot\mbox{\boldmath{$n$}}_{S}$. 
The result as presented on the RHS of Eq.~(\ref{e39}) is in full conformity 
with the RHS of Eq.~(\ref{e27}).

Through the above exercise we have determined the set of assumptions which
are needed in order to obtain the Kac-Moody algebra in Eq.~(\ref{e28})
in $d=2$. We now consider in how far our reliance upon `common wisdom' 
is justified. An aspect that immediately is noticed is the fact that 
${\cal O}(\vert\Delta\vert^3)$ on the RHS of Eq.~(\ref{e37}) is 
proportional to ${\rm d}^2 f(x)/{\rm d} x^2$ for {\sl some} $x$ in 
the open interval $(a,a+\Delta)$, implying that for a function $f(x)$ 
which is not twice continuously differentiable,
\footnote{\label{f23}
\small
To be explicit, the rest term on the RHS of Eq.~(\ref{e37}) is equal to 
$(-\Delta^3/12) f''(x)$, with $x\in (a,a+\Delta)$.} 
there exists {\sl no} reason, whatever, that the RHS of Eq.~(\ref{e37}) 
would be in any way related to the left-hand side. It is well-known that 
for FLs ${\sf n} ({\Bf k})$ is discontinuous at ${\Bf k}={\Bf k}_F$ and 
for LLs, ${\sf n}({\Bf k})$ though continuous at ${\Bf k}={\Bf k}_F$, 
is {\sl not} differentiable at this point. It is interesting, however, 
that independent of how one proceeds from Eq.~(\ref{e35}) to the final 
result, the term $\|{\Bf q}\| \cos\theta \equiv {\Bf q}\cdot 
\mbox{\boldmath{$n$}}_{S}$ remains unchanged so that the correct treatment 
of Eq.~(\ref{e36}) does {\sl not} lead to destruction of the Kac-Moody 
algebra. Nevertheless, here, contrary to the $d=1$ case, $\lambda$
{\sl cannot} be chosen arbitrarily large, but must satisfy the requirement
$\lambda\ll\Lambda$. As we shall demonstrate below, this restriction 
gives rise to a significant effect.

Owing to $\Lambda\ll \|{\Bf k}_F\|$, the restriction $\lambda\ll\Lambda$
justifies use of the following two approximations: $k_1(\theta) \equiv
\|{\Bf k}_F\| -\lambda/2 \approx \|{\Bf k}_F\|$ and $k_2(\theta) \equiv
\|{\Bf k}_F\| + \lambda/2 \approx \|{\Bf k}_F\|$. By the same reasoning
we can employ ${\sf n}\big({\Bf k}_1(\theta)\big)-{\sf n}
\big({\Bf k}_2(\theta)\big) \approx Z_{{\Bf k}_F}$. Further, since 
$k_2(\theta)-k_1(\theta) =\lambda$ and $0 \le {\sf n}({\Bf k}) \le 1$ 
(but ${\sf n}({\Bf k})\not\equiv 0$) for {\sl all} ${\Bf k}$, we have
\begin{equation}
\label{e1x}
0 < \int_{k_1(\theta)}^{k_2(\theta)} {\rm d}\|{\Bf k}\|\;
{\sf n}\big({\Bf k}(\theta)\big) < \lambda.
\end{equation}
Thus as long as $Z_{{\Bf k}_F} \gg \lambda/\|{\Bf k}_F\|$, we can neglect 
the integral on the RHS of Eq.~(\ref{e35}) and using the above
approximation for ${\sf n}\big({\Bf k}_1(\theta)\big)-{\sf n}
\big({\Bf k}_2(\theta)\big)$ write
\begin{equation}
\label{e2x}
g_S({\Bf q}) \approx \zeta^{-1}\, \Lambda\, {\Bf q}\cdot
\mbox{\boldmath{$n$}}_S,
\end{equation}
where $\zeta^{-1} {:=} Z_{{\Bf k}_F}$, to be compared with the expression 
in Eq.~(\ref{e39}). Consequently, the Kac-Moody algebra associated the 
$g_S({\Bf q})$ in Eq.~(\ref{e2x}) is similar to that in Eq.~(\ref{e28}), 
however with the RHS multiplied by $\zeta^{-1} \equiv Z_{{\Bf k}_F}$. 
Since ${\Bf q}$ is reciprocal to ${\Bf x}$, scaling of the RHS of 
Eq.~(\ref{e28}) by $\zeta^{-1}$ (here $Z_{{\Bf k}_F}$), gives rise to 
scaling of ${\Bf x}$, and by the requirement of the Lorentz invariance, 
of $t$, by $\zeta$. Thus with the RHS of Eq.~(\ref{e28}) being multiplied 
by $\zeta^{-1}$, we have (see Eq.~(\ref{e3b}) above)
\begin{equation}
\label{e39a}
G_{0;{\rm f}}({\Bf k},t;S) \rightharpoonup 
G_{0;{\rm f}}({\Bf k},t;S) + (1-\zeta) {\cal F}_{{\Bf k},\omega}\Big[ 
\frac{\Lambda}{(2\pi)^2}\,
\frac{\exp\big(i {\Bf k}_F(S)\cdot {\Bf x}\big)}{\zeta\,{\Bf x}\cdot
\mbox{\boldmath{$n$}}_S -\zeta\,t - i\,\eta\,{\rm sgn}(\zeta t)}
\Big],\;\eta\downarrow 0.
\end{equation}
It can be verified that this modification is `harmless' as far as the 
fundamental question with regard to {\sl nature} of the metallic
state of the system under consideration is concerned (i.e. whether
it is a FL or otherwise).

The situation undergoes a fundamental change when $Z_{{\Bf k}_F}$ is 
small, on the order of $\lambda/\|{\Bf k}_F\|$ or smaller, or vanishing, 
as is the case for such NFLs as marginal FLs and LLs. In such cases, the 
magnitude of the RHS of Eq.~(\ref{e35}), i.e. $\zeta^{-1} \|{\Bf k}_F\|$, 
is on the order of $\lambda$ (see Eq.~(\ref{e1x}) above) from which it 
follows that the associated Kac-Moody algebra is inadequate: since 
$\zeta^{-1}\approx \lambda/\|{\Bf k}_F\| \equiv 1/{\cal N}$ (see paragraph 
following Eq.~(\ref{e28}) above), in the case at hand the size of the 
quantum anomaly is by on the order of $1/\sqrt{{\cal N}}$ {\sl smaller} 
than that of `some error term' (see footnote \ref{f21}). 

It has now become evident why, as we have established above through 
explicit calculation (see Eq.~(\ref{e12}) and the following text), 
within the Haldane (1992) bosonisation scheme, the large-$\|{\Bf x}\|$ 
behaviour of $-i G({\Bf x},t=0^-)$ is always FL-like: the Kac-Moody 
algebra in Eq.~(\ref{e28}) is only compatible with metallic states 
whose corresponding $Z_{{\Bf k}_F} \gg \lambda/\|{\Bf k}_F\|$. This 
compatibility does not guarantee that Eq.~(\ref{e28}) gives a sufficiently
complete account of the physical processes at low energies of metallic 
systems for which holds $Z_{{\Bf k}_F} \gg \lambda/\|{\Bf k}_F\|$,
since our explicit calculations show that under {\sl all} 
circumstances $\Sigma(0)=0$, which is highly uncommon.

We note in passing that the result in Eq.~(\ref{e27}) is exact, 
for {\sl all} $d$, for the case where the coupling constant of the
particle-particle interaction is vanishing. From this perspective, our 
finding that within the framework of the Haldane (1992) bosonisation 
scheme $\Sigma(0)=0$ under {\sl all} conditions, is consistent with the 
fact that for non-interacting particles $\Sigma({\Bf k},\omega)$ is 
identically vanishing. 

In view of the above-indicated shortcoming of the Haldane bosonisation 
program in $d > 1$, we are presently not in a position to make {\sl any} 
well-founded judgement with regard to the true nature of the metallic 
state of the model under discussion. By taking into account the fact that 
an approximation for ${\bar\chi}_0(x)$, of the type presented in 
Eq.~(\ref{e16}), innocuous as it may seem at first glance, is capable 
of turning a ``FL'' into a ``LL'', we are of the opinion that the model 
considered here, may as yet have some surprising properties in stores, 
to be exposed by means of a future refined treatment.

% VI
\section{A brief discussion of some alternative bosonisation programmes}

In light of our finding with regard to a major shortcoming in the Haldane 
(1992) bosonisation scheme in $d > 1$, it is important to mention that in 
our considerations we have been dealing with a specific realisation of 
Haldane's scheme, namely one that has been explicitly dealt with by 
Haldane (1992) and Houghton and Marston (1993). In addition to this, two 
alternative realisations of this scheme have been put forward, one by 
Castro Neto and Fradkin (1994a,b, 1995) and the other by Kopietz and 
Sch\"onhammer (1996). 

In the realisation by Castro Neto and Fradkin (1994a,b, 1995), the authors 
construct a coherent-state path-integral formalism for the bosonised 
problem which concerns low-energy long-wavelength excitations of the 
fermionic problem. The coherent states in this approach are eigenstates 
of the annihilation operators $\{{\hat a}_{\Bf q}({\Bf k})\}$ which are 
constructed from the equal-time density operators and which under some 
assumptions, similar in spirit to those in the works by Haldane (1992) 
and Houghton and Marston (1993), the authors demonstrate to form a 
generalised Kac-Moody algebra. An aspect which is crucial to arriving at 
this result concerns replacing the density {\sl operator} in the momentum 
space with the {\sl expectation value} of it with respect to the ground 
state of the {\sl non-interacting} fermion system. We can demonstrate 
(Farid 1999b) that this approximation is at odds with a LL as well as 
some unconventional FL metallic states in $d > 1$, even though, as Castro 
Neto and Fradkin have shown (see Appendix B in Castro Neto and Fradkin 
1994b), it is compatible with the former in $d=1$; remarkably, the 
distinction between the two cases, corresponding to $d=1$ and $d > 1$, 
turns out to arise from the same mechanism that renders Eq.~(\ref{e28}) 
(or Eq.~(\ref{e27})) compatible with a LL metallic state in $d=1$ but 
incompatible in $d > 1$. 

In the approach by Kopietz and Sch\"onhammer (1996) (for a detailed
exposition see Kopietz 1997) the fermion problem is described in terms 
of an imaginary-time functional integral over Grassmann fields. The 
action in this description is subsequently related to a Hamiltonian 
which is subjected to Haldane's (1992) `patching' scheme of the FS. 
Following this, the two-body interaction is transformed away by means 
of a Hubbard-Stratonovich transformation. Through application of a 
second Hubbard-Stratonovich transformation, a composite Grassmann field 
is eliminated in the formalism in exchange for a collective bosonic 
field. The kinetic term of the effective action thus obtained, is 
subsequently dealt with through application of the perturbation theory. 
Within the framework of the Gaussian approximation, all terms beyond the 
second order in this kinetic contribution are neglected. Kopietz, 
Hermisson and Sch\"onhammer (1996) find justification for the Gaussian 
approximation in the so-called generalised closed-loop theorem (or 
generalised loop-cancellation theorem), applicable to all spatial 
dimensions 
\footnote{\label{f24}
\small For $d=1$ the loop-cancellation theorem is due to Dzyaloshinski\v{i} 
and Larkin (1974). For an extensive elaboration on this theorem, in $d=1$, 
see (Bohr 1981).}
$d$. The proof of this theorem in $d > 1$ is based upon two simplifying 
approximations, i) the so-called `diagonal-patch' approximation
\footnote{\label{f25}
\small
This is equivalent to the approximation as expressed through 
Eqs.~(\ref{e23}) and (\ref{e24}). }
and ii) local linearisation of the energy dispersion of the non-interacting 
fermions. For this theorem to be significant for applications, the effective 
screened interaction, in momentum space, amongst particles must be 
negligible for momenta ${\Bf q}$ whose magnitudes $\|{\Bf q}\|$ are larger 
than a cut-off $q_c$, with $q_c$ assumed to satisfy $q_c\ll\|{\Bf k}_F\|$. 
On the basis of the generalised closed-loop theorem, it has been shown that 
the just-mentioned Gaussian approximation becomes asymptotically exact for 
$q_c/\|{\Bf k}_F\|\to 0$. Validity of neither of the two mentioned 
conditions is dependent upon an equivalent of Eq.~(\ref{e28}). Since in 
the limit of long wavelengths and high density of fermions the theory 
based upon the Gaussian approximation has been shown to reduce to those 
by Houghton and Marston (1993) and Castro Neto and Fradkin (1994a,b) (see 
specifically \S~4.2.5 in Kopietz 1997), we may conclude that validity of 
the generalised closed-loop theorem in $d > 1$ must be restricted to FLs. 
That this may be the case can be made plausible (Farid 1999b) by 
considering the fact that validity of the generalised closed-loop theorem 
depends on the validity of the so-called `asymptotic velocity conservation' 
(see text following Eq.~(\ref{e39e}) below) in $d > 1$ (this `velocity 
conservation' is {\sl exact} in $d=1$ for much the same reason that 
Eq.~(\ref{e28}) is of general applicability in $d=1$ but {\sl not} in 
$d > 1$); for NFLs, since the Fermi velocity is by definition undefined, 
\footnote{\label{f26}
\small
This follows the fact that in metallic NFLs, $\Sigma({\Bf k},\omega_F)$ 
and/or $\Sigma({\Bf k}_F,\omega)$ fail, by definition, to be continuously 
differentiable in neighbourhoods of ${\Bf k}={\Bf k}_F$ and $\omega=
\omega_F$, respectively (see text preceding Eq.~(\ref{e1}) above).}  
one encounters a fundamental difficulty in rigorously demonstrating the
asymptotic validity of the generalised closed-loop theorem for cases 
where the metallic state is a NFL. The exactness of the loop-cancellation 
theorem in $d=1$, irrespective of the nature of the metallic state of 
the system, is directly related to existence of {\sl exact} (as opposed 
to asymptotically-correct) conservation laws, which embody the well-known 
Ward identities (see, e.g., Bohr 1981). To illustrate our line of 
reasoning, we mention that the essence of the `asymptotic velocity 
conversation' is expressed in the following result (see, e.g., Metzner, 
Castellani and Di Castro 1998):
\begin{eqnarray}
\label{e39d}
& &G_0({\Bf p}-{\Bf q}/2,\omega-\omega_0/2)\, 
G_0({\Bf p}+{\Bf q}/2,\omega+\omega_0/2)\nonumber\\
& &\;\;\;\;\;\;\;\;\;\;\;\;\;\;
= \frac{G_0({\Bf p}-{\Bf q}/2,\omega-\omega_0/2)
- G_0({\Bf p}+{\Bf q}/2,\omega+\omega_0/2)}{\omega_0
-{\Bf q}\cdot {\Bf v}^0_{\Bf p} + {\cal O}(\|{\Bf q}\|^m)},
\end{eqnarray}
where ${\Bf v}^0_{\Bf p} {:=} {\Bf\nabla}_{\Bf p} \omega^0_{\Bf p}$
and $m$ stands for an {\sl integer larger than unity}; for {\sl linear} 
dispersions, ${\cal O}(\|{\Bf q}\|^m)$ is vanishing. It can be trivially 
shown that
\begin{eqnarray}
\label{e39e}
& &G({\Bf p}-{\Bf q}/2,\omega-\omega_0/2)\, 
G({\Bf p}+{\Bf q}/2,\omega+\omega_0/2) \nonumber\\
& &\;\;\;\;\;\;\;\;\;\;\;\;\;\;
= \frac{G({\Bf p}-{\Bf q}/2,\omega-\omega_0/2)
- G({\Bf p}+{\Bf q}/2,\omega+\omega_0/2)}{\omega_0
-{\Bf q}\cdot {\Bf v}^0_{\Bf p} 
-[\Sigma({\Bf p}+{\Bf q}/2,\omega+\omega_0/2)
-\Sigma({\Bf p}-{\Bf q}/2,\omega-\omega_0/2)] 
+ {\cal O}(\|{\Bf q}\|^m)}.
\end{eqnarray}
The `asymptotic velocity conservation' to which we have referred above,
reflects the fact that denominator of the expression on the RHS of 
Eq.~(\ref{e39d}) asymptotically behaves like $(\omega_0-{\Bf q}\cdot 
{\Bf v}^0_{\Bf p})$ for $\|{\Bf q}\|\to 0$. This can be shown to be 
the case also for the denominator of the expression on the RHS of 
Eq.~(\ref{e39e}) {\sl provided} the system under consideration is a FL; 
for NFLs, $[\Sigma({\Bf p}+{\Bf q}/2,\omega+\omega_0/2)-\Sigma({\Bf p}
-{\Bf q}/2,\omega-\omega_0/2)]$ is the dominant contribution as compared 
with $(\omega_0 - {\Bf q}\cdot {\Bf v}_{\Bf p}^0)$ and therefore for NFLs 
the RHS of Eq.~(\ref{e39e}) {\sl cannot} be brought into a form resembling 
the RHS of Eq.~(\ref{e39d}). This amounts to the fact that the route to 
demonstrating the asymptotic validity of the closed-loop theorem on the 
basis of a perturbation series in terms of the {\sl interacting} Green 
function is closed when the system under consideration is not a FL, 
supporting the point of view that implications of this theorem may not 
be self-consistent. It is important to note that ${\Bf q}\cdot 
{\Bf v}^0_{\Bf p}$ in denominators of the expressions on the RHSs of 
Eqs.~(\ref{e39d}) and (\ref{e39e}) are of the same origin as ${\Bf q}\cdot
\mbox{\boldmath{$n$}}_S$ on the RHS of Eq.~(\ref{e28}) --- recall that 
since FS `patches' are flat, ${\Bf v}^{0}_{\Bf p}$, with ${\Bf p}\in S$, 
is parallel to $\mbox{\boldmath{$n$}}_S$. Note also that dominance of 
$[\Sigma({\Bf p}+{\Bf q}/2,\omega+\omega_0/2)-\Sigma({\Bf p}-{\Bf q}/2,
\omega-\omega_0/2)]$ in comparison with $(\omega_0-{\Bf q}\cdot 
{\Bf v}^0_{\Bf p})$, in cases of NFLs, is related to that of the 
suppressed terms on the RHS of Eq.~(\ref{e28}) when NFLs are concerned.

For completeness, we mention that Kopietz and Castilla (1996) have 
introduced a method of incorporating the non-linear corrections to the 
non-interacting energy dispersions and Kopietz (1997) has generalised 
the original formalism through which corrections beyond the Gaussian 
approximation can be accounted for.

% VII
\section{Summary and concluding remarks}

In this work we have first employed a method of characterising metallic 
states of systems of interacting fermions as FLs (including conventional 
and un-conventional ones) and NFLs in terms of the asymptotic behaviour 
of $\delta G(\omega)$ for small $\omega$. Following a close examination 
of some results in the energy-momentum domain obtained by earlier authors 
concerning a model of interacting fermions in two spatial dimensions, which 
proved inconclusive as regards the true metallic state of this model, we 
have studied (full details to be published elsewhere) some specific aspects 
of the single-particle Green function in the space-time domain. This study 
has confronted us with an inconsistency (namely that under {\sl all} 
conditions metallic states of systems in $d > 1$ must be unconventional 
FLs) which we have subsequently traced back to a fundamental shortcoming 
in an operator algebra that stands central to the process of bosonisation 
of the low-energy degrees of freedom of metallic systems of fermions in 
$d > 1$ spatial dimensions. We have explicitly shown that this shortcoming 
is due to inapplicability in $d > 1$ of an underlying assumption when 
NFLs, specifically LLs, are concerned.
\footnote{\label{f27}
\small
A careful study of the work by Mattis and Lieb (1965) reveals a remarkable 
parallel between the general inapplicability of the Haldane (1992) 
bosonisation method in $d > 1$ and inadequacy of the Luttinger's (1963) 
original treatment of the Luttinger model. }
On the other hand, since {\sl all} metallic states according to the
Haldane bosonisation scheme turn out to be unconventional FLs, it
follows that even in the case of FLs the mentioned operator algebra
does not describe the underlying physical processes sufficiently
accurately. It {\sl may} turn out that the mentioned shortcoming can be 
removed without needing to destroy the fundamental structure of the 
operator algebra, i.e. the Kac-Moody algebra, to which we have just 
referred. From these observations we have arrived at the conclusion that 
on the basis of the available results {\sl nothing} can be inferred as 
regards the true metallic state of the two-dimensional model system under 
consideration. 
$\hfill\Box$ 

\vspace{0.33cm}
\section*{Acknowledgements}
\noindent
It is a pleasure for me to thank Professor Peter B. Littlewood and 
members of the Theory of Condensed Matter Group for their kind 
hospitality at Cavendish Laboratory where this work was carried out.
I record my indebtedness to Girton College, Cambridge, for invaluable
support. I extend my thanks to Dr Peter Kopietz for kindly providing
me with references (Kopietz 1997) and (Bohr 1981). I dedicate this 
work to the memory of Dame Mary Cartwright (1900 - 1998).

%---------------------------------------------

%\vfill
%\pagebreak


\begin{references}
\centerline{\sc References}
\begin{verse}

{\sc Abramowitz}, M., and {\sc Stegun}, I.~A. (editors), 1972, 
{\sl Handbook of Mathematical Functions} (New York: Dover).\\

{\sc Anderson}, P.~W., 
1988, in {\sl Frontiers and Borderlines in Many-Particle Physics},
J.~R. Schrieffer and R.~A. Broglia (editors) (Amsterdam: North Holland), 
pp. 1-40;\\
--- 1989, in {\sl Strong Correlation and Superconductivity}
H. Fukuyama, S. Maekawa and A.~P. Malozemoff (editors) (Berlin: Springer 
Verlag), pp. 2-13;\\
--- 1990a, {\sl Phys. Rev. Lett.}, {\bf 64}, 1839;\\
--- 1990b, {\sl Phys. Rev. Lett.}, {\bf 65}, 2306;\\
--- 1991a, {\sl Physica} C, {\bf 185}-{\bf 198}, 11;\\
--- 1991b, {\sl Phys. Rev. Lett.}, {\bf 67}, 660;\\
--- 1992, {\sl Science}, {\bf 256}, 1526;\\
--- 1993, {\sl Phys. Rev. Lett.}, {\bf 71}, 1220;\\
--- 1997, {\sl THE Theory of Superconductivity in the High-$T_c$ 
Cuprate Superconductors} (Princeton, NJ: Princeton University Press).\\

{\sc Anderson}, P.~W., and {\sc Ren}, Y., 1990, in {\sl High Temperature 
Superconductivity}, K.~S. Bedell, D. Coffey, D.~E. Meltzer, D. Pines 
and J.~R. Schrieffer (editors), (Redwood City: Addison-Wesley).\\

{\sc Bares}, P.-A., and {\sc Wen}, X.-G., 1993, {\sl Phys. Rev.} B, 
{\bf 48}, 8636.\\

{\sc Bloom}, P., 1975, {\sl Phys. Rev.} B, {\bf 12}, 125.\\

{\sc Bohr}, T., 1981, {\sl The Luttinger Model} Lecture notes in the 
series on {\sl Low-Dimensional Statistical Mechanics} (Demark: {\sc 
nordita} preprint, {\sc nordita}-81/4).\\

{\sc Brink}, L., and {\sc Hanneaux}, M., 1988, {\sl Principles of 
String Theory} (New York: Plenum).\\

{\sc Castro Neto}, A.~H., and {\sc Fradkin}, E., 
1994a, {\sl Phys. Rev. Lett.}, {\bf 72}, 1393;\\
--- 1994b, {\sl Phys. Rev.} B, {\bf 49}, 10877;\\
--- 1995, {\sl Phys. Rev.} B, {\bf 51}, 4084.\\

{\sc Chakravarty}, S., {\sc Sudb{\o}}, A., {\sc Anderson}, P.~W.,
{\sc Strong}, S., 1993, {\sl Science}, {\bf 261}, 337.\\

{\sc DuBois}, D.~F., 1959, {\sl Ann. of Phys.}, {\bf 8}, 24.\\

{\sc Dzyaloshinski\v{i}}, I.~E., and {\sc Larkin}, A.~I., 1974,
{\sl Soviet Phys.} --- {\sl JETP}, {\bf 38}, 202.\\

{\sc Farid}, B., 1999a, {\sl Phil. Mag.} B, {\bf 79}, 1097;\\
--- 1999b to be published.\\

{\sc Fetter}, A.~L., and {\sc Walecka}, J.~D., 1971, {\sl Quantum 
Theory of Many-Particle Systems} (New York: McGraw-Hill).\\

{\sc Fujimoto}, S., 1990, {\sl J. Phys. Soc. Jpn}, {\bf 59}, 2316.\\

{\sc Fukuyama}, H., {\sc Narikiyo}, O., and {\sc Hasegawa}, Y.,
1991, {\sl J. Phys. Soc. Jpn}, {\bf 60}, 372.\\

{\sc Galitskii}, V.~M., 1958, {\sl Soviet Phys.} --- {\sl JETP},
{\bf 7}, 104.\\

{\sc Gell-Mann}, M., and {\sc Brueckner}, K., 1957, {\sl Phys. Rev.} B,
{\bf 106}, 364.\\

{\sc Goddard}, P., and {\sc Olive}, D., 1986, {\sl Int. J. Mod. Phys.},
{\bf 1}, 303.\\

{\sc Haldane}, F.~D.~M., 1980, {\sl Phys. Rev. Lett.}, {\bf 45}, 1358;\\
--- 1981, {\sl J. Phys.} C, {\bf 14}, 2585.\\
--- 1992, Proceedings of the International School of Physics 
``Enrico Fermi'', Course 121, J.~R. Schrieffer and R.~A. Broglia 
(editors), (New York: North Holland). 
For an abstract of some elements discussed in this work see: 
{\sc Haldane}, F.~D.~M., 1992, {\sl Helv. Phys. Acta}, {\bf 65}, 152.\\

{\sc Hodges}, C., {\sc Smith}, H., and {\sc Wilkins}, J.~W., 1971,
{\sl Phys. Rev.} B, {\bf 4}, 302.\\

{\sc Houghton}, A., {\sc Kwon}, H.-J., and {\sc Marston}, J.~B., 
1994, {\sl Phys. Rev.} B, {\bf 50}, 1351.\\

{\sc Houghton}, A., {\sc Kwon}, H.-J., {\sc Marston}, J.~B.,
and {\sc Shankar}, R., 1994, {\sl J. Phys.} C, {\bf 6}, 4909.\\

{\sc Houghton}, A., and {\sc Marston}, J.~B., 1993, {\sl Phys. Rev.} B,
{\bf 48}, 7790.\\

{\sc Hugenholtz}, N.~M., 1957, {\sl Physica}, {\bf 23}, 533 
(see p.~544).\\

{\sc Kopietz}, P., {\sc Hermisson}, J., and {\sc Sch\"onhammer}, K.,
1995, {\sl Phys. Rev.} B, {\bf 52}, 10877.\\

{\sc Kopietz}, P., and {\sc Castilla}, G.~E., 1996, {\sl Phys. Rev. 
Lett.}, {\bf 76}, 4777.\\

{\sc Kopietz}, P., and {\sc Sch\"onhammer}, K., 1996, {\sl Z. Phys.} B,
{\bf 100}, 259. [In pre-print form, July 1994.]\\

{\sc Kopietz}, P., 1997, {\sl Bosonization of Interacting Fermions 
in Arbitrary Dimensions} in {\sl Lecture Notes in Physics.
New Series m: Monographs} (Berlin: Springer).\\

{\sc Kwon}, H.-J., {\sc Houghton}, A., and {\sc Marston}, J.~B., 1995,
{\sl Phys. Rev.} B, {\bf 52}, 8002.\\

{\sc Lauwerier}, H.~A., 1977, {\sl Asymptotic Analysis}, Part I, 2nd
printing (Amsterdam: Mathematisch Centrum).\\

{\sc Littlewood}, P.~B., and {\sc Varma}, C.~M., 1991, {\sl J. Appl. 
Phys.}, {\bf 69}, 4979.\\

{\sc Luttinger}, J.~M., 1960, {\sl Phys. Rev.}, {\bf 119}, 1153;\\
--- 1961, {\sl Phys. Rev.}, {\bf 121}, 942;\\
--- 1963, {\sl J. Math. Phys.}, {\bf 4}, 1154.\\

{\sc Luttinger}, J.~M., and {\sc Ward}, J.~C., 1960, {\sl Phys. Rev.},
{\bf 118}, 1417.\\

{\sc Mahan}, G.~D., 1981, {\sl Many-Particle Physics} (New York: Plenum).\\

{\sc Mattis}, D.~C., and {\sc Lieb}, E.~H., 1965, {\sl J. Math. Phys.},
{\bf 6}, 304.\\

{\sc Metzner}, W., {\sc Castellani}, C., and {\sc Di Castro}, C.,
1998, {\sl Adv. Phys.}, {\bf 47}, 317.\\

{\sc Moroni}, S., {\sc Senatore}, G., and {\sc Fantoni}, S., 1997,
{\sl Phys. Rev.} B, {\bf 55}, 1040.\\ 

{\sc Murray}, J.~D., 1974, {\sl Asymptotic analysis} (Oxford: Oxford 
University Press).\\

{\sc Seitz}, F., 1940, {\sl Modern Theory of Solids} (New York: 
McGraw-Hill).\\

{\sc Varma}, C.~M., {\sc Littlewood}, P.~B., {\sc Schmitt-Rink}, S.,
{\sc Abrahams}, E., and {\sc Ruckenstein}, A.~E., 1989, {\sl Phys. 
Rev. Lett.}, {\bf 63}, 1996;\\
--- 1990 (E), {\sl Phys. Rev. Lett.}, {\bf 64}, 497.\\

{\sc Voit}, J., 1994, {\sl Rep. Prog. Phys.}, {\bf 57}, 977.\\

{\sc Wheatley}, J., {\sc Hsu}, T., and {\sc Anderson}, P.~W.,
1988a, {\sl Phys. Rev.} B, {\bf 37}, 5897;\\
--- 1988b, {\sl Nature}, {\bf 333}, 121.\\

{\sc Whittaker}, E.~T., and {\sc Watson}, G.~N., 1927, 4th edition 
(reprinted 1984) {\sl A Course of Modern Analysis} (Cambridge: 
Cambridge University Press).\\

\end{verse}
\end{references}
\end{document}